\documentclass[12pt]{article}

\usepackage{scicite}
\usepackage{times}

\newenvironment{sciabstract}{%
\begin{quote} \bf}
{\end{quote}}

\usepackage{geometry}

\usepackage{graphicx}%
\usepackage{tabularx}%
\usepackage{multirow}%
\usepackage{amsmath,amssymb,amsfonts}%
\usepackage{amsthm}%
\usepackage{mathrsfs}%
\usepackage[title]{appendix}%
\usepackage{xcolor}%
\usepackage{textcomp}%
\usepackage{manyfoot}%
\usepackage{booktabs}%
\usepackage{algorithm}%
\usepackage{algorithmicx}%
\usepackage{algpseudocode}%
\usepackage{listings}%
\usepackage{subcaption}%

\title{Understanding Discrepancies of Wavefunction Theories for Large Molecules}

\author
{Tobias Sch\"afer,$^{1\ast}$ Andreas Irmler,$^{1\ast}$, Alejandro Gallo,$^{1}$ Andreas Gr\"uneis$^{1\ast}$\\
\\
\normalsize{$^{1}$Institute for Theoretical Physics. TU Wien}\\
\normalsize{Wiedner Hauptstra\ss{}e 8--10/136, Vienna, A-1040, Austria}\\
\\
\normalsize{$^\ast$Corresponding authors. E-mail:  tobias.schaefer@tuwien.ac.at}
}

\date{}

\begin{document}

\maketitle

\begin{sciabstract}
Quantum mechanical many-electron calculations can predict properties of atoms, molecules and even complex materials.
The employed computational methods play a quintessential role in many scientifically and technologically relevant research fields.
However, a question of paramount importance is whether approximations aimed at reducing the computational complexity for solving the many-electron Schr\"odinger equation, are accurate enough.
Here, we investigate recently reported discrepancies of noncovalent interaction energies for large molecules predicted by two of the most widely-trusted many-electron theories:
diffusion quantum Monte Carlo and coupled-cluster theory. 
We are able to unequivocally pin down the source of the puzzling discrepancies and present  modifications to widely-used coupled-cluster methods needed for more accurate noncovalent interaction energies of large molecules on the hundred-atom scale.
This enhances the reliability of predictions from quantum mechanical many-electron theories across a wide range of critical applications, including drug design, catalysis, and the innovation of new functional materials, such as those for renewable energy technologies.
\end{sciabstract}

\section{Introduction}\label{sec1}

The prediction of electronic transition energies for a single hydrogen atom in 1926
marks the beginning of an incredible successful era of quantum mechanics~\cite{Schroedinger1926,Pauli1926}.
Shortly after this breakthrough, Paul Dirac famously noted that the underlying physical laws necessary for much of physics and all of chemistry are now completely known. 
However, he also pointed out that the exact application of these laws leads to equations that are much too complicated to solve~\cite{Dirac1929}.
This paradigm of theoretical chemistry and physics is prevalent until today.
In particular, the exponential growth of the computational complexity of the many-electron problem with system size
still makes an exact solution of the electronic Schr\"odinger equation for
more than a few atoms impossible.
As a consequence, a hierarchy of increasingly accurate methods
that is capable of producing reference results at the expense of tractable yet high computational cost has emerged.
These reference results are pivotal in order to
develop, assess, and further improve computationally more efficient but in general less accurate approximations.
In this context, a prime example was the numerical prediction of highly accurate ground state energies
for the uniform electron gas using diffusion Monte Carlo methods, leveraging the development
of approximate exchange-correlation functionals that ultimately led to the
breakthrough of density functional theory in computational materials science during the last decades~\cite{Kohn1965,Ceperley1980,Jones2015}.

At present, quantum mechanical many-electron calculations of systems containing more than 100 atoms
have become possible thanks to methodological developments and considerable growth in computing power.
These methodological improvements are often based on
taking advantage of the relative short-rangedness of many-electron correlation effects ~\cite{Riplinger2013,Qianli2019,Szabo2023}.
In this manner, the scaling of the computational complexity with respect to system size can be lowered.
However, recently several works~\cite{Benali2020,Hamdani2021,Ballesteros2021,Villot2022}
showed that there exist alarming discrepancies between predicted
interaction energies for large molecules when using two of the most widely-trusted highly accurate
many-electron theories: DMC and CCSD(T), which stand for diffusion Monte Carlo and coupled-cluster theory using single, double, and perturbative triple particle-hole excitation operators, respectively.
These observations are a source of great concern in the electronic structure theory community
because, in the case of noncovalent interactions between molecular complexes,
both CCSD(T)  and DMC are considered highly reliable benchmark methods~\cite{Rezac2016,Dubecky2016}.
Furthermore, the observed discrepancies are large enough to cause qualitative differences in calculated properties of materials, which can have scientific, technological, and even clinical implications.
For example, accurate crystal structure predictions are crucial in drug design to differentiate between harmful and effective polymorphs~\cite{Firaha2023,Hoja2019,Reilly2016}.
Similarly, reliable reference methods are essential for discovering and designing new functional materials for applications such as renewable energy storage and conversion, including catalysis, or solar cells~\cite{Chanussot2021,Sauer2024,Bokdam2017}.
Finally, as machine learning increasingly pervades all areas of computational first-principles physics, the accuracy of these reference methods, which provide the training data, becomes even more critical~\cite{Donchev2021,Eastman2023,Liu2023}.

In the following, we analyze a set of large molecular systems where large
discrepancies between approximated versions of DMC and CCSD(T) were
observed~\cite{Benali2020,Hamdani2021}.
Importantly, a direct experimental measurement of the computed interaction energies
of these systems is complicated and prone
to significant uncertainties.
Therefore, it is an open challenge to identify the origin of the observed deviations
for the employed highly accurate yet approximate theoretical approaches.

\section{Results}\label{sec2}

We present an approach which allows us to unambiguously test if the employed approximations for DMC and CCSD(T) cause the puzzling discrepancies between their predictions.
In particular, our methodology exhibits three striking advantages.
Firstly, due to its efficient and massive computational parallelization, we
omit any local correlation approximation, as was employed for the CCSD(T)
calculations in Refs.~\cite{Hamdani2021,Ballesteros2021,Villot2022}.
Secondly, we use a plane wave basis set to enable an unbiased assessment of
the quality of previously employed tabulated atom-centered basis functions.
Thirdly, we are able to study the influence of higher-order contributions
to the many-electron perturbation expansion beyond CCSD(T) theory for large molecular complexes.


In order to demonstrate the reliability of our plane wave basis approach, we first investigate the parallel displaced benzene dimer as a benchmark.
We find that our approach effectively addresses the challenges of noncovalent interactions between large molecules, combining the compactness and systematic improvability of natural orbitals without near-linear dependencies that plague atom-centered Gaussian basis sets for densely packed structures.
As discussed in the supplementary information, our computed CCSD(T) interaction energies for the parallel displaced benzene dimer are in excellent agreement with Gaussian basis set results.
Next, we turn to the coronene dimer interaction energy, where significant discrepancies between DMC and CCSD(T) have been observed~\cite{Benali2020,Hamdani2021}.
As shown in Tab.~\ref{tab:c2c2pd}, our canonical CCSD(T) estimates for the parallel displaced coronene dimer align closely with domain-based local pair-natural orbital (DLPNO-CCSD(T)) and local natural orbital (LNO-CCSD(T)) results, ruling out basis set incompleteness and local approximation errors as sources of discrepancies with DMC findings.
However, it is noteworthy that the CCSD(T) interaction energy contains a large (T) contribution of about
$-8\,\text{kcal/mol}$, indicating that the correct treatment of triple particle-hole excitation effects
for the electronic correlation plays a crucial role.

\begin{table}[]
    \centering
    \caption{%
      \textbf{Interaction energy in kcal/mol of the parallel displaced coronene dimer (C2C2PD) obtained at different
      levels of theory including MP2, CCSD, CCSD(T), CCSD(cT) and
      DMC.}  
      The uncertainty of the referenced DMC, LNO and DLPNO results are taken from the corresponding reference.
      The uncertainty of this work's results are dominated by the remaining basis set error and the uncertainty of the box size extrapolation. }
    \begin{tabularx}{\textwidth}{XXX}
      \toprule
      Theory & Interaction energy & Ref. \\
      \midrule
      MP2 & -38.5 $\pm$ 0.5 & this work \\
      CCSD & -13.4 $\pm$ 0.5 & this work \\
      CCSD(T) & -21.1 $\pm$ 0.5 & this work \\
      \midrule
      LNO-CCSD(T) & -20.6 $\pm$ 0.6       & Ref.~\cite{Hamdani2021} \\
      DLPNO-CCSD(T$_0$) & -20.9 $\pm$ 0.4 & Ref.~\cite{Villot2022} \\
      DMC & -18.1(8) & Ref.~\cite{Hamdani2021} \\
      DMC & -17.5(14) & Ref.~\cite{Benali2020} \\
      \midrule
      CCSD(cT) & -19.3 $\pm$ 0.5 & this work \\
      \bottomrule
    \end{tabularx}
    \label{tab:c2c2pd}
\end{table}


\subsection{All that glitters is not gold: overcorrelation in CCSD(T)}

Having ruled out errors from local approximations and incomplete basis sets for the parallel displaced coronene dimer (C2C2PD),
we seek to assess the (T) approximation, which makes a significant contribution to the interaction energy of C2C2PD.
In passing we anticipate that the (T) contribution to the interaction energy is also relatively large for
all other systems with a significant discrepancy reported in Ref.~\cite{Hamdani2021} (see supplementary information).

The (T) approximation was introduced in the seminal work by Raghavachari~\textit{et. al.}~\cite{Raghavachari1989}.
Since then, it has become one of the most widely-used benchmark methods
--sometimes referred to as the 'gold standard' of molecular quantum chemistry--
for weakly correlated systems.
However, we argue that the partly significantly too strong interaction energies
in CCSD(T) theory are caused by the
employed truncation of the approximation of the triple
particle-hole excitation operator.
These shortcomings are comparable to the issue of too strong interaction energies from truncated perturbation theories for systems with large polarizabilities, as discussed by Nguyen~\textit{et. al.}~\cite{Nguyen202    0}.
As can be observed for the coronene dimer in Table~\ref{tab:c2c2pd}, second-order M{\o}ller-Plesset perturbation theory (MP2)---a truncated pertubation theory---exhibits this overestimation of the interaction energy.
In the extreme case of an infinite polarizability, as it occurs in metallic systems,
MP2 and CCSD(T) even yield divergent correlation energies in the thermodynamic limit, which is referred to as infrared catastrophe~\cite{Shepherd2013,Masios2023}.
In contrast, a resummation of certain terms to infinite order can yield interaction energies with an accuracy that is
less dependent on the polarizability.
Prominent examples for such approaches include the CCSD theory
as well as the random-phase approximation. 
We have recently presented a method, denoted as CCSD(cT), that averts the infrared catastrophe of CCSD(T)
by including selected higher-order terms in the triples amplitude approximation without
significantly increasing the computational complexity~\cite{Masios2023}.

\medskip
\noindent\emph{Understanding the discrepancy}
\medskip

\noindent
For the present work it is important to note that the main difference
between CCSD(cT) and CCSD(T) theory originates from the employed approximation to the triple
particle-hole excitation amplitudes.
The triple amplitudes of the (cT) approximation are given in diagrammatic and algebraic form by~\cite{Masios2023}
\begin{equation}
  \includegraphics[width=0.8\textwidth]{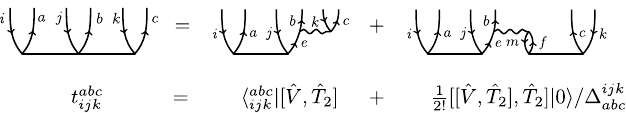}
  \label{eq:amplitude}
\end{equation}
where $\hat{V}$ and $\hat{T}_2$ stand for the Coulomb interaction and the double
particle-hole excitation operator, respectively. For brevity, the contributions from the single
excitation operator are not included and only one additional `direct' diagram is depicted.
In here, $\Delta^{ijk}_{abc}=\varepsilon_i+\varepsilon_j+\varepsilon_k-\varepsilon_a-\varepsilon_b-\varepsilon_c$, with
$\varepsilon$'s being one-electron HF energies. The bra- and ket-states
correspond to a triple excited and reference state, respectively.
The (T) approximation disregards the term $[[\hat{V},\hat{T}_2],\hat{T}_2]$, which
is included in (cT) and also occurs in full CCSDT theory.
This term effectively screens the bare
Coulomb interaction of the $[\hat{V},\hat{T}_2]$ term and has an opposite sign,
making it crucial for systems with large polarizability.
However, for small and weakly polarizible systems the $[[\hat{V},\hat{T}_2],\hat{T}_2]$ contribution is small,
making the (T) and (cT) approximation agree, as it
was already shown for a set of small molecules~\cite{Masios2023}.


We now demonstrate that using CCSD(cT) instead of CCSD(T) theory restores excellent agreement for
noncovalent interaction energies with DMC findings.
First, we consider again the coronene dimer. 
Table~\ref{tab:c2c2pd} shows that the binding energy for the coronene dimer calculated on the level of
CCSD(cT) theory is by almost $2\,\text{kcal/mol}$ closer to the DMC estimate compared to CCSD(T) theory,
achieving chemical accuracy ($1\,\text{kcal/mol}$) in comparison to DMC after subtracting error bars.
Next, we investigate the accuracy of CCSD(T) and CCSD(cT) compared to DMC 
for noncovalent interactions in smaller molecules. To this end, we study a set of dimers containing up to 24 atoms
that were also investigated in Ref.~\cite{Hamdani2021}.
This gives us another opportunity to assess the effect of local approximations at the level of CCSD(T) theory.
\begin{figure}
\includegraphics[]{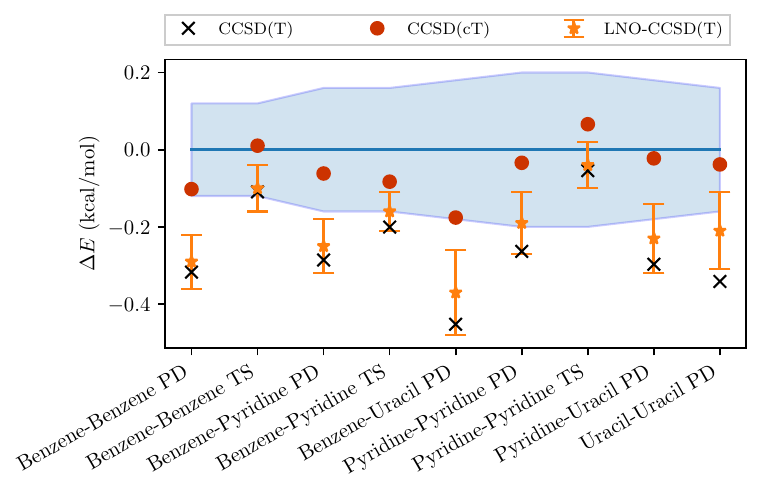}
\caption{%
  \textbf{Deviations of coupled cluster results from DMC results for a set of noncovalently bound dimers with up to 24 atoms.}
  The CCSD(T) and CCSD(cT) values are CBS estimates obtained from basis set extrapolation using aug-cc-pVTZ and aug-cc-pVQZ basis sets~\cite{Dunning1989,Kendall1992} (details in the supplementary information).
  The LNO-CCSD(T) and DMC data are from Ref.~\cite{Hamdani2021}.
  The uncertainty measures are as described in Table~\ref{tab:c2c2pd}.
  The uncertainty of DMC is shown by the blue area.
}
\label{fig:puzzle-systems-deviations}
\end{figure}
Fig.~\ref{fig:puzzle-systems-deviations} depicts the deviations of all computed interaction energies
from DMC reference values taken from Ref.~\cite{Hamdani2021}.
It should be noted that DMC references and differences to LNO-CCSD(T) interaction energies are
shown with error bars~\cite{Hamdani2021}.
Using our massive computational parallelization approach, we are able to add
canonical CCSD(T) interaction energies extrapolated to the CBS limit
to the comparison to DMC.\label{text:hpcgauss}
For these relatively small molecules, we can employ sufficiently large basis sets, reducing the remaining uncertainty to approximately 0.01 kcal/mol (see supplementary information).
Importantly, our canonical CCSD(T) results are in good agreement with LNO-CCSD(T) findings to within its error
bars. The only minor exception is observed for the parallel displaced uracil dimer, where canonical CCSD(T)
predicts a slightly stronger interaction.
A comparison to DMC reveals that CCSD(T) theory predicts on average about $0.3\,\text{kcal/mol}$
stronger interaction energies. Based on LNO-CCSD(T) and DMC data alone such a statement cannot be made due to
the relatively large and mostly overlapping error bars. However, our well converged canonical CCSD(T) findings allow
drawing such conclusions.
Only for the T-shaped pyridine and benzene dimers,
DMC and CCSD(T) binding energies agree to within the DMC errors. Note that these systems have a smaller (T) contribution to the intereaction energy, compared to the parallel displaced systems. All other systems exhibit small but significant discrepancies
between CCSD(T) and DMC results, which is consistent with the even larger discrepancies reported for the
larger molecules in Ref.~\cite{Hamdani2021}.
Similar to our findings for the coronene dimer reported in Table~\ref{tab:c2c2pd},
Fig.~\ref{fig:puzzle-systems-deviations} shows that CCSD(cT) interaction energies
agree significantly better with DMC values than their CCSD(T) counterparts.
\begin{figure}
  \centering
  \includegraphics[]{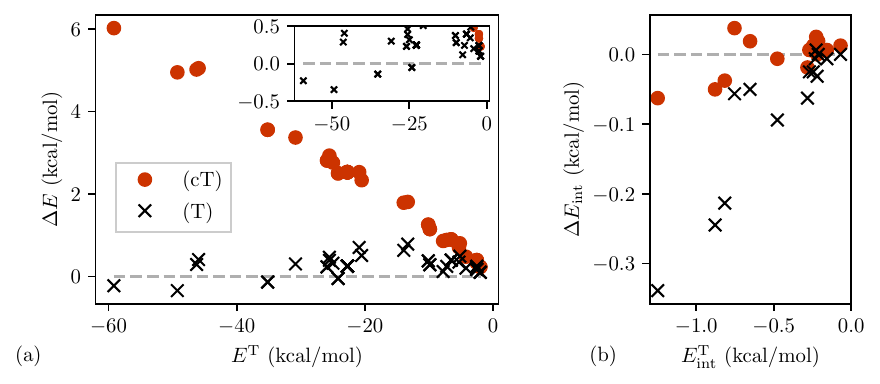}
  \caption{%
    \textbf{Comparison between the full triples  and the perturbative triples approaches, (cT) and (T) for a set of molecules contained in the S22 data set~\cite{S22}.}
    The total triples correlation energy contribution $E^T$ on the x-axis is compared to
    both differences between the (T), (cT) correlation energy contributions and $E^T$ for
    \textbf{a)} total energies and
    \textbf{b)} interaction energies.
    \label{fig:s22}}
\end{figure}

Given the good agreement between DMC and CCSD(cT) for the systems studied above, an important question to ask is if CCSD(cT) is really more accurate than CCSD(T) for noncovalent interaction energies?
To answer this question we compare interaction energies of both approaches to higher-level CC methods for complexes from the S22 data set. 
As can be observed in Fig. \ref{fig:s22}, we find that while (T) is in good agreement with T for total energies, it overestimates interaction energies.
Here T stands for the triples contribution to the correlation energy, $E^\text T = E^{\text{CCSDT}}-E^{\text{CCSD}}$.
(cT) closely matches the T interaction energies, indicating its superior accuracy for weakly bound complexes.
This effect is particularly strong for interaction energies with large triples correlation contributions.


\subsection{Estimating the overcorrelation of (T) for weak interactions}

In summary, we have demonstrated that CCSD(cT) theory achieves excellent agreement for noncovalent interaction energies between molecular complexes
compared to DMC and CCSDT theory.
However, we stress that the CC series of methods (CCSD, CCSDT and CCSDTQ) is observed to yield monotonic and rapidly converging
interaction energies for small and weakly bound complexes~\cite{Smith2014}.
Based on this knowledge, we emphasize that the Q contribution to the interaction energies
can be expected to be smaller than its T counterpart, but could possibly yield a significant contribution.
Indeed, this is part of the reason for the success of the CCSD(T) approximation for very small molecules,
where CCSD(T) is often fortuitously closer to CCSDTQ than CCSDT~\cite{Smith2014}.
Here, we argue that this error cancellation no longer functions in the case of large molecular complexes
involving strongly polarizable systems such as C2C2PD, C3GC and C60@[6]CPPA.
A similar problem is known to occur in MP2 theory, where the truncation of the perturbation series
also leads to significantly too strong interaction energies for systems with large polarizability,
although MP2 yields relatively accurate
interaction energies for systems with an intermediate polarizability~\cite{Nguyen2020}.
\begin{figure}
  \centering
  \includegraphics[]{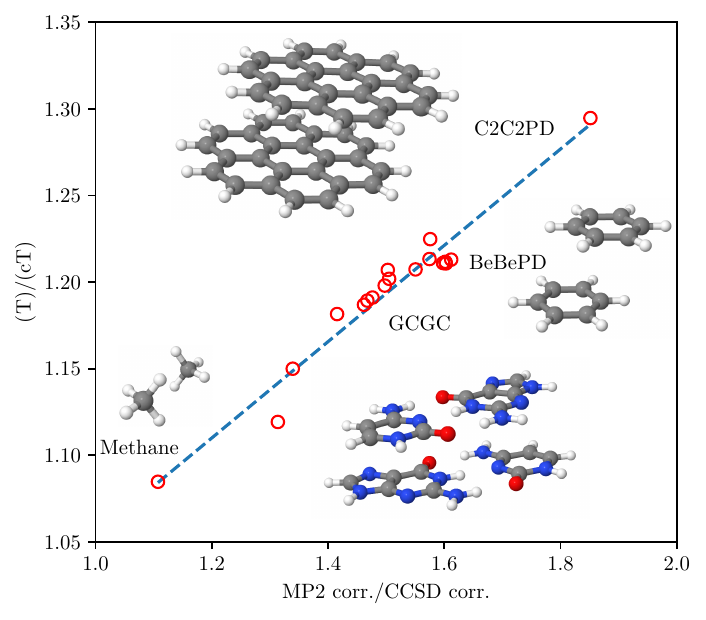}
  \caption{%
    \textbf{Correlation between the ratio of (T) and (cT) with the ratio of the MP2 and CCSD correlation energy contributions to interaction energies of a set of dispersion-dominated complexes from the S22, L7 and S66 benchmark datasets~\cite{S22, L7, S66}.}
    Selected cases are labeled and visualized: Methane dimer, GCGC (guanine-cytosine tetramer), BeBePD (benzene dimer parallel displaced), C2C2PD (coronene dimer parallel displaced).
   All the data is available in the supplementary information.
  }
  \label{fig:ratios}
\end{figure}
To quantify and support the statements above,
Fig.~\ref{fig:ratios} illustrates that there exists a correlation between the ratio of (T) and (cT) 
with the ratio of the MP2 and CCSD correlation energy contributions to the interaction energies
of all studied molecules in this work with dispersion-dominated interactions.
This demonstrates that (T) exhibits a tendency to overestimate the absolute binding energy in a
similar manner as MP2 theory for more polarizable systems.
Although an overestimation of the (T) binding energy contribution compared to its (cT) counterpart
by about 10\% might yield a fortuitously better agreement between CCSD(T) and CCSDTQ,
we argue that 20--30\% overestimation is expected to yield significantly too strong interaction energies.
For example, the values of (T)/(cT) for the Benzene-Benzene PD and coronene dimer are
1.2 and 1.3, respectively.

Having demonstrated and explained the reasons for the overestimation of absolute interaction
energies on the level of CCSD(T) theory for small molecules with up to 24 atoms and the C2C2PD system,
we now want to turn to the discussion of the remaining large molecular complexes where
significant absolute discrepancies between DMC and CCSD(T) have been observed.
These systems include C3GC from the L7 data set and the C60@[6]CPPA buckyball-ring.
Here, substantial differences in the binding energies of $2.2\,\text{kcal/mol}$ and  $7.6\,\text{kcal/mol}$
were reported, respectively after subtracting error bars.
Although CCSD(cT) calculations for systems of that size are currently not feasible using
our approach, we now introduce a simplified model that allows us to estimate the change
in interaction energies from CCSD(T) to CCSD(cT) in an approximate manner.
Given the linear trend between the different correlation energy contributions to the interaction energy
depicted in Fig.~\ref{fig:ratios},
it is possible to estimate the (cT) contribution for systems where only MP2, CCSD and (T) are known.
These numbers can be calculated using a computationally efficient LNO-CCSD(T) implementation~\cite{MRCC}.
Results computed in this manner are denoted as CCSD(cT)-fit.
Details about this procedure and the corresponding error estimates are provided in the supplementary information.
%
\begin{table}[t]
  \caption{%
     \textbf{Comparison of the interaction energy for large molecular complexes in kcal/mol as calculated by different levels of theory.}
     Showcasing partially large discrepancies between CCSD(T) and DMC on the one hand, and an excellent agreement between
     CCSD(cT) and DMC results for complexes up to the 100-atom scale on the other hand.
     CCSD(T) and CCSD(cT) results are obtained using our plane wave approach.
     The calculation and the uncertainty of CCSD(cT)-fit is explained in the supplementary information. 
    \label{tab:dmc-vs-pt}}
\resizebox{\textwidth}{!}{%
\begin{tabular}{lcccccc}
\toprule
System    & CCSD(T)             & LNO-CCSD(T)~\cite{Hamdani2021} & CCSD(cT) & CCSD(cT)-fit & DMC~\cite{Hamdani2021} & DMC~\cite{Benali2020} \\
\midrule
GGG       &   -1.5 $\pm$ 0.5    &   -2.1 $\pm$ 0.2  &   -1.2 $\pm$ 0.5 &  -1.8 $\pm$ 0.2  &  -1.5(6)  &  -2.0(8)  \\
GCGC      &  -13.1 $\pm$ 0.5    &  -13.6 $\pm$ 0.4  &  -12.5 $\pm$ 0.5 & -12.8 $\pm$ 0.5  & -12.4(8)  & -10.6(12) \\
C2C2PD    &  -21.1 $\pm$ 0.5    &  -20.6 $\pm$ 0.6  &  -19.3 $\pm$ 0.5 & -18.9 $\pm$ 0.7  & -18.1(8)  & -17.5(14) \\
C3A       &                     &  -16.5 $\pm$ 0.8  &                  & -15.3 $\pm$ 0.9  & -15.0(10) & -16.6(18) \\
PHE       &                     &  -25.4 $\pm$ 0.2  &                  & -25.0 $\pm$ 0.2  & -26.5(13) & -24.9(12) \\
C3GC      &                     &  -28.7 $\pm$ 1.0  &                  & -26.7 $\pm$ 1.1  & -24.2(13) & -25.1(18) \\
$\text C_{60}$@[6]CPPA &        &  -41.7 $\pm$ 1.7  &                  & -35.6 $\pm$ 2.0  & -31.1(14) &  \\
\bottomrule
\end{tabular}}
\end{table}
Table~\ref{tab:dmc-vs-pt} gives our estimated CCSD(cT) interaction energies
in comparison to CCSD(T) and DMC findings 
for seven large molecular complexes.
A comparison between CCSD(cT)-fit and the explicitly calculated CCSD(cT) results for GGG, GCGC and C2C2PD shows
that the linear regression model is sufficiently reliable for the systems studied in this work.
For comparison Table~\ref{tab:dmc-vs-pt} also summarizes the DMC interaction energies
from Refs.~\cite{Benali2020} and~\cite{Hamdani2021}, which agree to within
at least $1\,\text{kcal/mol}$ for GGG, C2C2PD, PHE and C3GC.
For the remaining systems the DMC estimates show a larger discrepancy and for C60@[6]CPPA only one DMC estimate is available.
Although the DMC binding energies have overlapping error bars, the remaining uncertainties are relatively large,
illustrating that obtaining highly accurate interaction energies for these large molecules
is also challenging for DMC.

As already discussed in Ref.~\cite{Hamdani2021}, CCSD(T) interaction energies listed Table~\ref{tab:dmc-vs-pt} 
exhibit large discrepancies compared to DMC for
C2C2PD, C3GC and C60@[6]CPPA.
In contrast, CCSD(cT)-fit resolves these discrepancies
for all systems on the hundred-atom scale, achieving excellent agreement with
DMC estimates of Hamdani~\textit{et. al.}~\cite{Hamdani2021}
to within chemical accuracy ($1\,\text{kcal/mol}$) after subtracting the error bars.
Even for C60@[6]CPPA, which contains 132 atoms, a discrepancy of only $1.1\,\text{kcal/mol}$ remains,
although the error bar of CCSD(cT)-fit is relatively large in this case.
We argue that the remaining discrepancies
are potentially caused by uncertainties in DMC, CCSD(cT)-fit and the underlying LNO-CCSD(T)
calculations.
It should be noted that
the error bars of LNO-CCSD(T) interaction energies are in some cases underestimated,
as exemplified for the Uracil-Uracil PD dimer by the comparison between canonical CCSD(T) and LNO-CCSD(T) interaction energies
shown in Fig.~\ref{fig:puzzle-systems-deviations}.
Furthermore, the DMC interaction energy of C60@[6]CPPA has not yet been verfied independently using a different DMC implementation
as it was done for all other systems listed in Table~\ref{tab:dmc-vs-pt}.
We also stress that in some cases the differences between the DMC estimates are larger than their respective error bars.

\section{Conclusion}

Our work unequivocally demonstrates that,
due to the employed truncation of the many-body perturbation series expansion,
one of the most widely-used and accurate quantum chemistry approaches
-- CCSD(T) theory -- in certain cases binds noncovalently interacting
large molecular complexes too strongly. 
Our findings show that a simple yet efficient modification denoted as CCSD(cT) remedies these shortcomings.
This paves the way for highly reliable benchmark calculations
of large molecular complexes on the hundred-atom scale that play a crucial role in
scientific and technological problems, for example, drug design and surface science.
We stress that the more accurate CCSD(cT) approximation can directly be transferred to
computationally efficient low-scaling and local correlation approaches, which will substantially
advance the applications of theoretical chemistry as well as physics
in all areas of computational materials science where
highly accurate benchmark results are urgently needed.
We are witnessing an unremitting expansion of the frontiers of accurate
electronic structure theories to ever larger systems which when combined
with machine-learning techniques, has the potential to transform the paradigm
of modern computational materials science.

\bibliography{refs}

\begin{thebibliography}{10}

\bibitem{Schroedinger1926}
E.~Schr\"odinger, {\it Annalen der Physik\/} {\bf 384}, 361 (1926).

\bibitem{Pauli1926}
W.~{Pauli}, {\it Zeitschrift fur Physik\/} {\bf 36}, 336 (1926).

\bibitem{Dirac1929}
P.~A.~M. {Dirac}, {\it Proceedings of the Royal Society of London Series A\/}
  {\bf 123}, 714 (1929).

\bibitem{Kohn1965}
W.~Kohn, L.~J. Sham, {\it Phys. Rev.\/} {\bf 140}, A1133 (1965).

\bibitem{Ceperley1980}
D.~M. Ceperley, B.~J. Alder, {\it Phys. Rev. Lett.\/} {\bf 45}, 566 (1980).

\bibitem{Jones2015}
R.~O. Jones, {\it Rev. Mod. Phys.\/} {\bf 87}, 897 (2015).

\bibitem{Riplinger2013}
C.~Riplinger, B.~Sandhoefer, A.~Hansen, F.~Neese, {\it The Journal of chemical
  physics\/} {\bf 139} (2013).

\bibitem{Qianli2019}
Q.~Ma, H.-J. Werner, {\it Journal of Chemical Theory and Computation\/} {\bf
  15}, 1044 (2019).

\bibitem{Szabo2023}
P.~B. Szabó, J.~Csóka, M.~Kállay, P.~R. Nagy, {\it Journal of Chemical
  Theory and Computation\/} {\bf 19}, 8166 (2023).

\bibitem{Benali2020}
A.~Benali, H.~Shin, O.~Heinonen, {\it The Journal of Chemical Physics\/} {\bf
  153}, 194113 (2020).

\bibitem{Hamdani2021}
Y.~S. Al-Hamdani, {\it et~al.\/}, {\it Nature Communications\/} {\bf 12}, 3927
  (2021).

\bibitem{Ballesteros2021}
F.~Ballesteros, S.~Dunivan, K.~U. Lao, {\it The Journal of Chemical Physics\/}
  {\bf 154}, 154104 (2021).

\bibitem{Villot2022}
C.~Villot, F.~Ballesteros, D.~Wang, K.~U. Lao, {\it The Journal of Physical
  Chemistry A\/} {\bf 126}, 4326 (2022).

\bibitem{Rezac2016}
J.~Řezáč, P.~Hobza, {\it Chemical Reviews\/} {\bf 116}, 5038 (2016).

\bibitem{Dubecky2016}
M.~Dubecký, L.~Mitas, P.~Jurečka, {\it Chemical Reviews\/} {\bf 116}, 5188
  (2016).

\bibitem{Firaha2023}
D.~Firaha, {\it et~al.\/}, {\it Nature\/} {\bf 623}, 324 (2023).

\bibitem{Hoja2019}
J.~Hoja, {\it et~al.\/}, {\it Science Advances\/} {\bf 5}, eaau3338 (2019).

\bibitem{Reilly2016}
A.~M. Reilly, {\it et~al.\/}, {\it Acta Crystallographica Section B: Structural
  Science, Crystal Engineering and Materials\/} {\bf 72}, 439 (2016).

\bibitem{Chanussot2021}
L.~Chanussot, {\it et~al.\/}, {\it ACS Catalysis\/} {\bf 11}, 6059 (2021).

\bibitem{Sauer2024}
J.~Sauer, {\it Journal of Catalysis\/} {\bf 433}, 115482 (2024).

\bibitem{Bokdam2017}
M.~Bokdam, J.~Lahnsteiner, B.~Ramberger, T.~Schäfer, G.~Kresse, {\it Physical
  Review Letters\/} {\bf 119}, 145501 (2017).

\bibitem{Donchev2021}
A.~G. Donchev, {\it et~al.\/}, {\it Scientific Data 2021 8:1\/} {\bf 8}, 1
  (2021).

\bibitem{Eastman2023}
P.~Eastman, {\it et~al.\/}, {\it Scientific Data 2022 10:1\/} {\bf 10}, 1
  (2023).

\bibitem{Liu2023}
P.~Liu, {\it et~al.\/}, {\it Physical Review Letters\/} {\bf 130}, 078001
  (2023).

\bibitem{Raghavachari1989}
K.~Raghavachari, G.~W. Trucks, J.~A. Pople, M.~Head-Gordon, {\it Chemical
  Physics Letters\/} {\bf 157}, 479 (1989).

\bibitem{Nguyen2020}
B.~D. Nguyen, {\it et~al.\/}, {\it Journal of Chemical Theory and
  Computation\/} {\bf 16}, 2258 (2020).

\bibitem{Shepherd2013}
J.~J. Shepherd, A.~Gr\"uneis, {\it Phys. Rev. Lett.\/} {\bf 110}, 226401
  (2013).

\bibitem{Masios2023}
N.~Masios, A.~Irmler, T.~Sch\"afer, A.~Gr\"uneis, {\it Phys. Rev. Lett.\/} {\bf
  131}, 186401 (2023).

\bibitem{Dunning1989}
J.~Dunning, Thom~H., {\it The Journal of Chemical Physics\/} {\bf 90}, 1007
  (1989).

\bibitem{Kendall1992}
R.~A. Kendall, T.~H. Dunning, R.~J. Harrison, {\it J. Chem. Phys.\/} {\bf 96},
  6796 (1992).

\bibitem{S22}
P.~Jure\v{c}ka, J.~\v{S}poner, J.~\v{C}ern\'{y}, P.~Hobza, {\it Phys. Chem.
  Chem. Phys.\/} {\bf 8}, 1985 (2006).

\bibitem{Smith2014}
D.~G.~A. Smith, P.~Jankowski, M.~Slawik, H.~A. Witek, K.~Patkowski, {\it
  Journal of Chemical Theory and Computation\/} {\bf 10}, 3140 (2014).

\bibitem{L7}
R.~Sedlak, {\it et~al.\/}, {\it Journal of Chemical Theory and Computation\/}
  {\bf 9}, 3364 (2013).

\bibitem{S66}
J.~Řez\'a{}\v{c}, K.~E. Riley, P.~Hobza, {\it Journal of Chemical Theory and
  Computation\/} {\bf 7}, 2427 (2011).

\bibitem{MRCC}
M.~K\'a{}llay, {\it et~al.\/}, {\it The Journal of Chemical Physics\/} {\bf
  152}, 074107 (2020).

\end{thebibliography}


\begin{thebibliography}{11}%
\makeatletter
\providecommand \@ifxundefined [1]{%
 \@ifx{#1\undefined}
}%
\providecommand \@ifnum [1]{%
 \ifnum #1\expandafter \@firstoftwo
 \else \expandafter \@secondoftwo
 \fi
}%
\providecommand \@ifx [1]{%
 \ifx #1\expandafter \@firstoftwo
 \else \expandafter \@secondoftwo
 \fi
}%
\providecommand \natexlab [1]{#1}%
\providecommand \enquote  [1]{``#1''}%
\providecommand \bibnamefont  [1]{#1}%
\providecommand \bibfnamefont [1]{#1}%
\providecommand \citenamefont [1]{#1}%
\providecommand \href@noop [0]{\@secondoftwo}%
\providecommand \href [0]{\begingroup \@sanitize@url \@href}%
\providecommand \@href[1]{\@@startlink{#1}\@@href}%
\providecommand \@@href[1]{\endgroup#1\@@endlink}%
\providecommand \@sanitize@url [0]{\catcode `\\12\catcode `\$12\catcode
  `\&12\catcode `\#12\catcode `\^12\catcode `\_12\catcode `\%12\relax}%
\providecommand \@@startlink[1]{}%
\providecommand \@@endlink[0]{}%
\providecommand \url  [0]{\begingroup\@sanitize@url \@url }%
\providecommand \@url [1]{\endgroup\@href {#1}{\urlprefix }}%
\providecommand \urlprefix  [0]{URL }%
\providecommand \Eprint [0]{\href }%
\providecommand \doibase [0]{http://dx.doi.org/}%
\providecommand \selectlanguage [0]{\@gobble}%
\providecommand \bibinfo  [0]{\@secondoftwo}%
\providecommand \bibfield  [0]{\@secondoftwo}%
\providecommand \translation [1]{[#1]}%
\providecommand \BibitemOpen [0]{}%
\providecommand \bibitemStop [0]{}%
\providecommand \bibitemNoStop [0]{.\EOS\space}%
\providecommand \EOS [0]{\spacefactor3000\relax}%
\providecommand \BibitemShut  [1]{\csname bibitem#1\endcsname}%
\let\auto@bib@innerbib\@empty
\bibitem [{\citenamefont {K\'a{}llay}\ \emph {et~al.}(2020)\citenamefont
  {K\'a{}llay}, \citenamefont {Nagy}, \citenamefont {Mester}, \citenamefont
  {Rolik}, \citenamefont {Samu}, \citenamefont {Csontos}, \citenamefont
  {Cs\'o{}ka}, \citenamefont {Szab\'o{}}, \citenamefont {Gyevi-Nagy},
  \citenamefont {H\'e{}gely}, \citenamefont {Ladj\'a{}nszki}, \citenamefont
  {Szegedy}, \citenamefont {Lad\'o{}czki}, \citenamefont {Petrov},
  \citenamefont {Farkas}, \citenamefont {Mezei},\ and\ \citenamefont
  {Ganyecz}}]{MRCC}%
  \BibitemOpen
  \bibfield  {author} {\bibinfo {author} {\bibfnamefont {M.}~\bibnamefont
  {K\'a{}llay}}, \bibinfo {author} {\bibfnamefont {P.~R.}\ \bibnamefont
  {Nagy}}, \bibinfo {author} {\bibfnamefont {D.}~\bibnamefont {Mester}},
  \bibinfo {author} {\bibfnamefont {Z.}~\bibnamefont {Rolik}}, \bibinfo
  {author} {\bibfnamefont {G.}~\bibnamefont {Samu}}, \bibinfo {author}
  {\bibfnamefont {J.}~\bibnamefont {Csontos}}, \bibinfo {author} {\bibfnamefont
  {J.}~\bibnamefont {Cs\'o{}ka}}, \bibinfo {author} {\bibfnamefont {P.~B.}\
  \bibnamefont {Szab\'o{}}}, \bibinfo {author} {\bibfnamefont {L.}~\bibnamefont
  {Gyevi-Nagy}}, \bibinfo {author} {\bibfnamefont {B.}~\bibnamefont
  {H\'e{}gely}}, \bibinfo {author} {\bibfnamefont {I.}~\bibnamefont
  {Ladj\'a{}nszki}}, \bibinfo {author} {\bibfnamefont {L.}~\bibnamefont
  {Szegedy}}, \bibinfo {author} {\bibfnamefont {B.}~\bibnamefont
  {Lad\'o{}czki}}, \bibinfo {author} {\bibfnamefont {K.}~\bibnamefont
  {Petrov}}, \bibinfo {author} {\bibfnamefont {M.}~\bibnamefont {Farkas}},
  \bibinfo {author} {\bibfnamefont {P.~D.}\ \bibnamefont {Mezei}}, \ and\
  \bibinfo {author} {\bibfnamefont {A.}~\bibnamefont {Ganyecz}},\ }\href
  {\doibase 10.1063/1.5142048} {\bibfield  {journal} {\bibinfo  {journal} {The
  Journal of Chemical Physics}\ }\textbf {\bibinfo {volume} {152}},\ \bibinfo
  {pages} {074107} (\bibinfo {year} {2020})}\BibitemShut {NoStop}%
\bibitem [{\citenamefont {\texttt{CC4S}~developer team}(2024)}]{cc4s-manual}%
  \BibitemOpen
  \bibfield  {author} {\bibinfo {author} {\bibnamefont {\texttt{CC4S}~developer
  team}},\ }\href {https://manuals.cc4s.org/user-manual/} {\enquote {\bibinfo
  {title} {\texttt{CC4S} user manual},}\ }\bibinfo {howpublished}
  {\url{https://manuals.cc4s.org/user-manual/}} (\bibinfo {year}
  {2024})\BibitemShut {NoStop}%
\bibitem [{\citenamefont {Péter R.~Nagy}\ and\ \citenamefont
  {Kállay}(2023)}]{nagy.s66.basis}%
  \BibitemOpen
  \bibfield  {author} {\bibinfo {author} {\bibfnamefont {B.~D.~L.}\
  \bibnamefont {Péter R.~Nagy}, \bibfnamefont {László Gyevi-Nagy}}\ and\
  \bibinfo {author} {\bibfnamefont {M.}~\bibnamefont {Kállay}},\ }\href
  {\doibase 10.1080/00268976.2022.2109526} {\bibfield  {journal} {\bibinfo
  {journal} {Molecular Physics}\ }\textbf {\bibinfo {volume} {121}},\ \bibinfo
  {pages} {e2109526} (\bibinfo {year} {2023})}\BibitemShut {NoStop}%
\bibitem [{\citenamefont {Valiev}\ \emph {et~al.}(2010)\citenamefont {Valiev},
  \citenamefont {Bylaska}, \citenamefont {Govind}, \citenamefont {Kowalski},
  \citenamefont {Straatsma}, \citenamefont {{Van Dam}}, \citenamefont {Wang},
  \citenamefont {Nieplocha}, \citenamefont {Apra}, \citenamefont {Windus},\
  and\ \citenamefont {{de Jong}}}]{Valiev2010}%
  \BibitemOpen
  \bibfield  {author} {\bibinfo {author} {\bibfnamefont {M.}~\bibnamefont
  {Valiev}}, \bibinfo {author} {\bibfnamefont {E.}~\bibnamefont {Bylaska}},
  \bibinfo {author} {\bibfnamefont {N.}~\bibnamefont {Govind}}, \bibinfo
  {author} {\bibfnamefont {K.}~\bibnamefont {Kowalski}}, \bibinfo {author}
  {\bibfnamefont {T.}~\bibnamefont {Straatsma}}, \bibinfo {author}
  {\bibfnamefont {H.}~\bibnamefont {{Van Dam}}}, \bibinfo {author}
  {\bibfnamefont {D.}~\bibnamefont {Wang}}, \bibinfo {author} {\bibfnamefont
  {J.}~\bibnamefont {Nieplocha}}, \bibinfo {author} {\bibfnamefont
  {E.}~\bibnamefont {Apra}}, \bibinfo {author} {\bibfnamefont {T.}~\bibnamefont
  {Windus}}, \ and\ \bibinfo {author} {\bibfnamefont {W.}~\bibnamefont {{de
  Jong}}},\ }\href {\doibase https://doi.org/10.1016/j.cpc.2010.04.018}
  {\bibfield  {journal} {\bibinfo  {journal} {Comput. Phys. Commun.}\ }\textbf
  {\bibinfo {volume} {181}},\ \bibinfo {pages} {1477 } (\bibinfo {year}
  {2010})}\BibitemShut {NoStop}%
\bibitem [{\citenamefont {Irmler}\ \emph {et~al.}(2021)\citenamefont {Irmler},
  \citenamefont {Gallo},\ and\ \citenamefont {Gr{\"{u}}neis}}]{Irmler2021}%
  \BibitemOpen
  \bibfield  {author} {\bibinfo {author} {\bibfnamefont {A.}~\bibnamefont
  {Irmler}}, \bibinfo {author} {\bibfnamefont {A.}~\bibnamefont {Gallo}}, \
  and\ \bibinfo {author} {\bibfnamefont {A.}~\bibnamefont {Gr{\"{u}}neis}},\
  }\href {\doibase 10.1063/5.0050054} {\bibfield  {journal} {\bibinfo
  {journal} {The Journal of Chemical Physics}\ }\textbf {\bibinfo {volume}
  {154}},\ \bibinfo {pages} {234103} (\bibinfo {year} {2021})},\ \Eprint
  {http://arxiv.org/abs/2103.06788} {2103.06788} \BibitemShut {NoStop}%
\bibitem [{\citenamefont {Kresse}\ and\ \citenamefont
  {Furthm\"uller}(1996)}]{Efficiency.of.aKresse1996}%
  \BibitemOpen
  \bibfield  {author} {\bibinfo {author} {\bibfnamefont {G.}~\bibnamefont
  {Kresse}}\ and\ \bibinfo {author} {\bibfnamefont {J.}~\bibnamefont
  {Furthm\"uller}},\ }\href {\doibase 10.1016/0927-0256(96)00008-0} {\bibfield
  {journal} {\bibinfo  {journal} {Computational Materials Science}\ }\textbf
  {\bibinfo {volume} {6}},\ \bibinfo {pages} {15} (\bibinfo {year}
  {1996})}\BibitemShut {NoStop}%
\bibitem [{\citenamefont {Gruber}\ \emph {et~al.}(2018)\citenamefont {Gruber},
  \citenamefont {Liao}, \citenamefont {Tsatsoulis}, \citenamefont {Hummel},\
  and\ \citenamefont {Gr\"uneis}}]{Gruber2018}%
  \BibitemOpen
  \bibfield  {author} {\bibinfo {author} {\bibfnamefont {T.}~\bibnamefont
  {Gruber}}, \bibinfo {author} {\bibfnamefont {K.}~\bibnamefont {Liao}},
  \bibinfo {author} {\bibfnamefont {T.}~\bibnamefont {Tsatsoulis}}, \bibinfo
  {author} {\bibfnamefont {F.}~\bibnamefont {Hummel}}, \ and\ \bibinfo {author}
  {\bibfnamefont {A.}~\bibnamefont {Gr\"uneis}},\ }\href {\doibase
  10.1103/PhysRevX.8.021043} {\bibfield  {journal} {\bibinfo  {journal} {Phys.
  Rev. X}\ }\textbf {\bibinfo {volume} {8}},\ \bibinfo {pages} {021043}
  (\bibinfo {year} {2018})}\BibitemShut {NoStop}%
\bibitem [{\citenamefont {Gr{\"{u}}neis}\ \emph {et~al.}(2011)\citenamefont
  {Gr{\"{u}}neis}, \citenamefont {Booth}, \citenamefont {Marsman},
  \citenamefont {Spencer}, \citenamefont {Alavi},\ and\ \citenamefont
  {Kresse}}]{Gruneis2011}%
  \BibitemOpen
  \bibfield  {author} {\bibinfo {author} {\bibfnamefont {A.}~\bibnamefont
  {Gr{\"{u}}neis}}, \bibinfo {author} {\bibfnamefont {G.~H.}\ \bibnamefont
  {Booth}}, \bibinfo {author} {\bibfnamefont {M.}~\bibnamefont {Marsman}},
  \bibinfo {author} {\bibfnamefont {J.}~\bibnamefont {Spencer}}, \bibinfo
  {author} {\bibfnamefont {A.}~\bibnamefont {Alavi}}, \ and\ \bibinfo {author}
  {\bibfnamefont {G.}~\bibnamefont {Kresse}},\ }\href {\doibase
  10.1021/ct200263g} {\bibfield  {journal} {\bibinfo  {journal} {J. Chem.
  Theory Comput.}\ }\textbf {\bibinfo {volume} {7}},\ \bibinfo {pages} {2780}
  (\bibinfo {year} {2011})}\BibitemShut {NoStop}%
\bibitem [{\citenamefont {Hummel}\ \emph {et~al.}(2017)\citenamefont {Hummel},
  \citenamefont {Tsatsoulis},\ and\ \citenamefont {Grüneis}}]{Hummel2017}%
  \BibitemOpen
  \bibfield  {author} {\bibinfo {author} {\bibfnamefont {F.}~\bibnamefont
  {Hummel}}, \bibinfo {author} {\bibfnamefont {T.}~\bibnamefont {Tsatsoulis}},
  \ and\ \bibinfo {author} {\bibfnamefont {A.}~\bibnamefont {Grüneis}},\
  }\href {\doibase 10.1063/1.4977994} {\bibfield  {journal} {\bibinfo
  {journal} {The Journal of Chemical Physics}\ }\textbf {\bibinfo {volume}
  {146}},\ \bibinfo {pages} {124105} (\bibinfo {year} {2017})}\BibitemShut
  {NoStop}%
\bibitem [{\citenamefont {Knizia}\ \emph {et~al.}(2009)\citenamefont {Knizia},
  \citenamefont {Adler},\ and\ \citenamefont {Werner}}]{Knizia2009}%
  \BibitemOpen
  \bibfield  {author} {\bibinfo {author} {\bibfnamefont {G.}~\bibnamefont
  {Knizia}}, \bibinfo {author} {\bibfnamefont {T.~B.}\ \bibnamefont {Adler}}, \
  and\ \bibinfo {author} {\bibfnamefont {H.~J.}\ \bibnamefont {Werner}},\
  }\href {\doibase 10.1063/1.3054300/908511} {\bibfield  {journal} {\bibinfo
  {journal} {Journal of Chemical Physics}\ }\textbf {\bibinfo {volume} {130}}
  (\bibinfo {year} {2009}),\ 10.1063/1.3054300/908511}\BibitemShut {NoStop}%
\bibitem [{\citenamefont {Al-Hamdani}\ \emph {et~al.}(2021)\citenamefont
  {Al-Hamdani}, \citenamefont {Nagy}, \citenamefont {Zen}, \citenamefont
  {Barton}, \citenamefont {K{\'a}llay}, \citenamefont {Brandenburg},\ and\
  \citenamefont {Tkatchenko}}]{Hamdani2021}%
  \BibitemOpen
  \bibfield  {author} {\bibinfo {author} {\bibfnamefont {Y.~S.}\ \bibnamefont
  {Al-Hamdani}}, \bibinfo {author} {\bibfnamefont {P.~R.}\ \bibnamefont
  {Nagy}}, \bibinfo {author} {\bibfnamefont {A.}~\bibnamefont {Zen}}, \bibinfo
  {author} {\bibfnamefont {D.}~\bibnamefont {Barton}}, \bibinfo {author}
  {\bibfnamefont {M.}~\bibnamefont {K{\'a}llay}}, \bibinfo {author}
  {\bibfnamefont {J.~G.}\ \bibnamefont {Brandenburg}}, \ and\ \bibinfo {author}
  {\bibfnamefont {A.}~\bibnamefont {Tkatchenko}},\ }\href {\doibase
  10.1038/s41467-021-24119-3} {\bibfield  {journal} {\bibinfo  {journal}
  {Nature Communications}\ }\textbf {\bibinfo {volume} {12}},\ \bibinfo {pages}
  {3927} (\bibinfo {year} {2021})}\BibitemShut {NoStop}%
\end{thebibliography}%
\bibliographystyle{Science}

\section*{Acknowledgments}
Tobias Sch\"afer acknowledges support from the Austrian Science Fund (FWF) [DOI: 10.55776/ESP335].
The computational results presented have been largely achieved using the Vienna Scientific Cluster (VSC).
Andreas Irmler and Alejandro Gallo acknowledge support from the European Union's Horizon 2020 research and innovation program under Grant Agreement No. 951786 (The NOMAD CoE).
Support and funding from the European Research Council (ERC) (Grant Agreement No. 101087184) is gratefully acknowledged.
We gratefully acknowledge discussions with Adrian Daniel Boese.


\section*{Supplementary materials}
Materials and Methods\\
Supplementary Text\\
Figs. S1 to S2\\
Tables S1 to S5\\

\end{document}


\title{Supplementary information for: Understanding Discrepancies of Wavefunction Theories for Large Molecules}
\author{Tobias Sch\"afer}
\email{tobias.schaefer@tuwien.ac.at}
\author{Andreas Irmler}
\author{Alejandro Gallo}
\author{Andreas Gr\"uneis}
\affiliation{
  Institute for Theoretical Physics, TU Wien,\\
  Wiedner Hauptstraße 8-10/136, 1040 Vienna, Austria
}

\maketitle
\tableofcontents

\setlength{\LTcapwidth}{\textwidth}

\renewcommand{\tablename}{Table S}
\renewcommand{\figurename}{Fig. S}
\renewcommand{\refname}{Supplementary References}
\def\thesection{S\arabic{section}}
\def\thesubsection{S(\arabic{subsection})}
\def\thesubsubsection{S(\arabic{subsubsection})}

\section{Supplementary information notes}

\begin{itemize}

\item The CBS estimates for benzene-benzene PD (used in Fig.~1) and the
Gaussian basis CCSD(T) estimate of -2.70~kcal/mol are presented in
section~\hyperref[sec:s66]{S2}. Furthermore, the CBS estimates for CCSD(T) and
CCSD(cT) shown in Fig.~2 can be found in
section~\hyperref[sec:s66]{S2}. These numbers are from counterpoise corrected
calculations from AVQZ Hartree--Fock calculations together with correlation
calculations using a [34] extrapolation. All values are provided in
Table~\hyperref[tab:mrcc-mp2]{S \RNum 1}
\item Results depicted in Fig.~3 for molecules contained in the S22 data set on the level of
T, (T) and (cT) theory are summarized in \hyperref[sec:s22]{S3}
\item The plane wave approach used for the results in Fig.~1, Table~1, Fig.~4,
and Table~2 is explained section~\hyperref[sec:planewave]{S4}
\item Results for (T) energy contributions of the large molecular systems
mentioned in Section 2.2 can be found in Table~\hyperref[tab:lno]{S \RNum 5}
\item The fitting procedure in Section 2.3 is explained in section~\hyperref[sec:fit]{S5}
\item Data used in Fig.~4 is summarized in Table~\hyperref[tab:ints]{S \RNum 4}

\end{itemize}

\section{Selected S66 systems - canonical CBS estimates}
\label{sec:s66}

Here we provide highly accurate CBS estimates using Gaussian type orbitals for
9 molecules from the S66 test set.
For Gaussian type orbitals there exists a well established strategy to reach
the complete basis set limit (CBS). We use Dunning's correlation consistent
basis sets of type aug-cc-pVXZ (AVXZ), where X refers to the cardinal number of the basis set.
In this work, we employ X = T, Q, and 5. CBS estimates in post-Hartree--Fock
methods are obtained by a two point extrapolation assuming a X$^{-3}$
convergence. The extrapolation using AVTZ and AVQZ is denoted as [34], whereas
[45] employs the basis sets AVQZ and AV5Z.

Table~\hyperref[tab:mrcc-mp2]{S \RNum 1} shows the convergence of the interaction energies of
the studied molecules with respect to the employed basis set. Results are given
for Hartree--Fock, as well as for MP2, CCSD, (T) and the (cT) contribution.
Correlation energies were obtained using the \texttt{MRCC}~\cite{MRCC}
interfaced to our \texttt{Cc4s} code~\cite{cc4s-manual}.

For these systems, we are able to calculate canonical MP2 results with the
AVTZ, AVQZ, and AV5Z basis. One can see that the counterpoise corrected (CP)
and the uncorrected (NC) results agree well when the largest available basis
set is used. For HF the CP and NC with the AV5Z deviate only by 0.009~kcal/mol
or 0.014~kcal/mol for root mean square deviation (rms) and maximal deviation
(max), respectively. For MP2 the best available estimate for the complete basis
set would be the [45] extrapolation.  Here CP and NC deviate by 0.076 and
0.117~kcal/mol rms and max, respectively.  It can be seen that results from
smaller basis sets are significantly better for CP then for the uncorrected
case.  For HF the CP corrected results using AVQZ and AV5Z are for all intents
and purposes identical, with a maximum deviation of 0.002~kcal/mol.  In the NC
case AVQZ and AV5Z differ by 0.064 and 0.109~kcal/mol for rms and max,
respectively.  The same can be observed for MP2, here for the CP results the
[34] result is already very close to the [45] value, namely 0.007 and
0.017~kcal/mol for rms and max, respectively.  NC results show a larger
deviation of 0.112 and 0.250~kcal/mol for rms and max, respectively.

These results allow to conclude that both, AVQZ for HF and [34] extrapolation
for MP2, are sufficiently accurate for the given set of systems.

Now we can turn to the CCSD(T) correlation energies. As the BSIE of CCSD(T) is
known to be equal and mostly even smaller than in MP2, the provided results
obtained from [34] extrapolation are expected to be very close to the CBS.
Consequently, the expected deviations from the CBS limit  are in the order of
0.01~kcal/mol or lower.
These findings are in accordance with CBS estimates from Nagy et al.~\cite{nagy.s66.basis}
for the same system using slightly smaller basis sets.

\renewcommand*{\arraystretch}{1.1}
\begin{longtable}[H]{l|c| ccccc| ccccc}
\caption{%
Interaction energies in kcal/mol of the studied molecular systems with
different Gaussian basis sets. Shown is the Hartree--Fock energy contribution
as well as the canonical correlation energies for MP2, CCSD, (T), and (cT).
Both counterpoise (CP) corrected results as well as results without CP are
presented.
}
\label{tab:mrcc-mp2}\\
\toprule
                    &        & \multicolumn{5}{c|}{CP corrected} & \multicolumn{5}{c}{CP uncorrected}\\
                    & Method & AVTZ     &  AVQZ    &  AV5Z    & [34]   &  [45]  &  AVTZ     &   AVQZ   &    AV5Z  & [34]   & [45]   \\
\midrule
\endhead
\bottomrule
\endfoot
Pyridine-pyridine PD & HF  &   3.336 &   3.332 &   3.331 &    -    &    -    &   3.112 &   3.273 &   3.324 &    -    &   -     \\
                     & MP2 corr.&  -9.100 &  -9.238 &  -9.287 &  -9.339 &  -9.339 & -10.203 &  -9.717 &  -9.503 &  -9.362 &  -9.279 \\
                     & CCSD corr.&  -5.673 &  -5.743 &    -    &  -5.794 &    -    &  -6.597 &  -6.061 &    -    &  -5.670 &    -    \\
                     & (T) &  -1.254 &  -1.286 &    -    &  -1.310 &    -    &  -1.341 &  -1.321 &    -    &  -1.307 &    -    \\
                     & (cT) &  -1.029 &  -1.059 &    -    &  -1.080 &    -    &  -1.114 &  -1.093 &   -    &  -1.078 &  -  \\
\hline
Pyridine-pyridine TS   & HF  &   0.869 &   0.867 &   0.867 &    -    &    -    &   0.703 &   0.822 &   0.861 &    -    &   -     \\
                     & MP2 corr. &  -5.054 &  -5.172 &  -5.213 &  -5.258 &  -5.255 &  -6.016 &  -5.581 &  -5.400 &  -5.263 &  -5.211 \\
                     & CCSD corr.&  -3.457 &  -3.538 &    -    &  -3.597 &    -    &  -4.267 &  -4.267 &    -    &  -3.479 &    -   \\
                     & (T) &  -0.727 &  -0.747 &    -    &  -0.762 &    -    &  -0.801 &  -0.776 &    -    &  -0.758 &    -   \\
                     & (cT) &  -0.609 &  -0.628 &    -    &  -0.642 &    -    &  -0.682 &  -0.657 &   -    &  -0.639 &  - \\
\hline
Benzene-pyridine PD    & HF  &   3.621 &   3.619 &   3.618 &    -    &    -    &   3.395 &   3.559 &   3.610 &    -    &   -     \\
                     & MP2 corr. &  -8.838 &  -8.962 &  -9.006 &  -9.052 &  -9.053 &  -9.991 &  -9.439 &  -9.223 &  -9.036 &  -8.996 \\
                     & CCSD corr.&  -5.552 &  -5.609 &    -    &  -5.651 &    -    &  -6.523 &  -5.925 &    -    &  -5.488 &    -    \\
                     & (T) &  -1.229 &  -1.260 &    -    &  -1.282 &    -    &  -1.320 &  -1.295 &    -    &  -1.277 &    -    \\
                     & (cT) &  -1.009 &  -1.038 &    -    &  -1.058 &    -    &  -1.098 &  -1.072 &   -    &  -1.054 &  - \\
\hline
Benzene-pyridine TS     & HF  &   0.943 &   0.943 &   0.943 &    -    &    -    &   0.745 &   0.896 &   0.936 &    -    &   -     \\
                     & MP2 corr. &  -4.936 &  -5.042 &  -5.079 &  -5.119 &  -5.119 &  -6.016 &  -5.473 &  -5.273 &  -5.077 &  -5.063 \\
                     & CCSD corr.&  -3.369 &  -3.438 &    -    &  -3.488 &    -    &  -4.288 &  -3.725 &    -    &  -3.315 &    -    \\
                     & (T) &  -0.702 &  -0.721 &    -    &  -0.735 &    -    &  -0.786 &  -0.752 &    -    &  -0.728 &    -    \\
                     & (cT) & -0.587 &  -0.605 &    -    &  -0.618 &    -    &  -0.668 &  -0.635 &   -    &  -0.611 &  -  \\

\hline
Pyridine-uracil PD    & HF  &   2.074 &   2.072 &   2.071 &    -    &    -    &   1.730 &   1.984 &   2.060 &    -    &   -     \\
                     & MP2 corr. & -10.358 & -10.556 & -10.631 & -10.701 & -10.710 & -11.935 & -11.262 & -10.945 & -10.770 & -10.613 \\
                     & CCSD corr.&  -6.097 &  -7.024 &    -    &  -7.110 &    -    &  -8.266 &  -7.516 &    -    &  -6.969 & - \\
                     & (T) &  -1.567 &  -1.607 &    -    &  -1.637 &    -    &  -1.692 &  -1.658 &    -    &  -1.634 & - \\
                     & (cT) &  -1.298 &  -1.335 &    -    &  -1.362 &    -    &  -1.421 &  -1.386 &   -    &  -1.360 &  - \\
\hline
Benzene-benzene PD      & HF  &   3.964 &   3.961 &   3.960 &    -    &    -    &   3.739 &   3.901 &   3.952 &    -    &   -     \\
                     & MP2 corr. &  -8.479 &  -8.587 &  -8.625 &  -8.665 &  -8.666 &  -9.654 &  -9.052 &  -8.837 &  -8.613 &  -8.610 \\
                     & CCSD corr.&  -5.350 &  -5.392 &    -    &  -5.423 &    -    &  -6.345 &  -5.699 &    -    &  -5.227 &    -    \\
                     & (T) &  -1.183 &  -1.212 &    -    &  -1.234 &    -    &  -1.276 &  -1.247 &    -    &  -1.226 &    -    \\
                     & (cT) &  -0.973 &  -0.999 &    -    &  -1.019 &    -    &  -1.063 &  -1.033 &   -    &  -1.012 &  -      \\
\hline
Benzene-benzene TS      & HF  &   1.448 &   1.447 &   1.448 &    -    &    -    &   1.251 &   1.400 &   1.440 &    -    &   -     \\
                     & MP2 corr. &  -5.030 &  -5.125 &  -5.158 &  -5.194 &  -5.193 &  -6.158 &  -5.554 &  -5.352 &  -5.114 &  -5.141 \\
                     & CCSD corr.&  -3.419 &  -3.476 &    -    &  -3.518 &    -    &  -4.385 &  -3.762 &    -    &  -3.308 &    -    \\
                     & (T) &  -0.715 &  -0.734 &    -    &  -0.748 &    -    &  -0.802 &  -0.765 &    -    &  -0.738 &    -    \\
                     & (cT) &  -0.598 &  -0.616 &    -    &  -0.628 &    -    &  -0.682 &  -0.646 &   -    &  -0.620 &  -      \\
\hline
Uracil-uracil PD        & HF  &   0.388 &   0.379 &   0.377 &    -    &    -    &  -0.091 &   0.254 &   0.363 &    -    &   -     \\
                     & MP2 corr. & -11.047 & -11.323 & -11.430 & -11.525 & -11.542 & -13.152 & -12.298 & -11.872 & -11.675 & -11.425 \\
                     & CCSD corr.&  -7.802 &  -7.998 &    -    &  -8.142 &    -    &  -9.660 &  -8.700 &    -    &  -8.000 &    -    \\
                     & (T) &  -1.887 &  -1.936 &    -    &  -1.972 &    -    &  -2.052 &  -2.006 &    -    &  -1.972 &    -    \\
                     & (cT) &  -1.590 &  -1.636 &    -    &  -1.669 &    -    &  -1.751 &  -1.704 &   -    &  -1.670 &  -      \\
\hline
Benzene-uracil PD       & HF  &   3.444 &   3.443 &   3.442 &    -    &    -    &   3.070 &   3.352 &   3.431 &    -    &   -     \\
                     & MP2 corr. & -10.658 & -10.845 & -10.914 & -10.982 & -10.987 & -12.348 & -11.566 & -11.232 & -10.996 & -10.882 \\
                     & CCSD corr.&  -7.158 &  -7.260 &    -    &  -7.334 &    -    &  -8.620 &  -7.759 &    -    &  -7.131 &    - \\
                     & (T) &  -1.598 &  -1.640 &    -    &  -1.670 &    -    &  -1.732 &  -1.692 &    -    &  -1.664 &    - \\
                     & (cT) &  -1.327 &  -1.366 &    -    &  -1.394 &    -    &  -1.457 &  -1.417 &   -    &  -1.389 &  -   \\
\end{longtable}

\newpage

\section{Selected S22 systems - CCSDT using cc-pVDZ basis sets}
\label{sec:s22}

Here we provide the numerical data for 15 out of 22 molecular systems from the
S22 benchmark set.  Results are shown for canonical calculations using the
cc-pVDZ basis set.  The HF calculations were performed using
\texttt{NWCHEM}~\cite{Valiev2010} and for the post-HF methods we employed an interface to our
\texttt{Cc4s} code~\cite{Irmler2021}.
Interaction energies shown in the manuscript are evaluated from data summarized in
Table~\ref{tab:s22} and are not counterpoise corrected.

\renewcommand*{\arraystretch}{1.1}
\begin{longtable}[H]{rlrrrrrrrrl}
\caption{%
Results for molecular systems of the S22 dataset using the cc-pVDZ basis set in Hartree units.
T correlation energy contribution is evaluated via CCSDT-CCSD.
The \textit{index} column denotes the identifier of the molecule in the S22 dataset.
The \textit{Type} column describes the type of system that has been computed,
F1, F2 being the separate fragments, whereas Full denotes the whole system.
} \\
 \toprule
Index & System & Type & HF & MP2 corr. & CCSD corr. & (T) & (cT) & T  \\
 \midrule
 \endhead
 \bottomrule
 \endfoot
1     & Ammonia dimer                        & F1   &     -56.195616 &      -0.186397 &      -0.202634 &      -0.003802 &      -0.003531 &      -0.004072 \\
      &                                      & F2   &     -56.195616 &      -0.186397 &      -0.202634 &      -0.003802 &      -0.003531 &      -0.004072 \\
      &                                      & Full &    -112.396243 &      -0.375576 &      -0.407333 &      -0.007953 &      -0.007378 &      -0.008498 \\ \hline
2     & Water dimer                          & F1   &     -76.026603 &      -0.201874 &      -0.211441 &      -0.003051 &      -0.002844 &      -0.003214 \\
      &                                      & F2   &     -76.026710 &      -0.201741 &      -0.211310 &      -0.003041 &      -0.002835 &      -0.003203 \\
      &                                      & Full &    -152.062536 &      -0.406176 &      -0.424477 &      -0.006434 &      -0.005987 &      -0.006755 \\ \hline
3     & Formicacid dimer                     & F1   &    -188.778390 &      -0.500591 &      -0.512818 &      -0.015288 &      -0.013906 &      -0.015736 \\
      &                                      & F2   &    -188.778390 &      -0.500591 &      -0.512818 &      -0.015288 &      -0.013906 &      -0.015736 \\
      &                                      & Full &    -377.586256 &      -1.005708 &      -1.028426 &      -0.031867 &      -0.028964 &      -0.032677 \\ \hline
4     & Formamide dimer                      & F1   &    -168.945898 &      -0.483914 &      -0.502205 &      -0.015525 &      -0.014134 &      -0.016126 \\
      &                                      & F2   &    -168.945898 &      -0.483914 &      -0.502205 &      -0.015525 &      -0.014134 &      -0.016126 \\
      &                                      & Full &    -337.916259 &      -0.972313 &      -1.007590 &      -0.032177 &      -0.029269 &      -0.033301 \\ \hline
8     & Methane dimer                        & F1   &     -40.198702 &      -0.161179 &      -0.184661 &      -0.003702 &      -0.003457 &      -0.004092 \\
      &                                      & F2   &     -40.198702 &      -0.161179 &      -0.184661 &      -0.003702 &      -0.003457 &      -0.004092 \\
      &                                      & Full &     -80.396906 &      -0.323344 &      -0.370213 &      -0.007506 &      -0.007007 &      -0.008291 \\ \hline
9     & Ethene dimer                         & F1   &     -78.039915 &      -0.274850 &      -0.305009 &      -0.009796 &      -0.008999 &      -0.010426 \\
      &                                      & F2   &     -78.039915 &      -0.274850 &      -0.305009 &      -0.009796 &      -0.008999 &      -0.010426 \\
      &                                      & Full &    -156.079311 &      -0.552564 &      -0.612197 &      -0.019981 &      -0.018347 &      -0.021231 \\ \hline
10    & Benzene-Methane complex              & F1   &    -230.722189 &      -0.782664 &      -0.822131 &      -0.035812 &      -0.032184 &      -0.036209 \\
      &                                      & F2   &     -40.198646 &      -0.161040 &      -0.184473 &      -0.003686 &      -0.003442 &      -0.004074 \\
      &                                      & Full &    -270.919640 &      -0.947566 &      -1.009489 &      -0.039973 &      -0.036041 &      -0.040713 \\ \hline
11    & Benzene dimer PD                     & F1   &    -230.722178 &      -0.782673 &      -0.822141 &      -0.035811 &      -0.032184 &      -0.036209 \\
      &                                      & F2   &    -230.722178 &      -0.782673 &      -0.822141 &      -0.035811 &      -0.032184 &      -0.036209 \\
      &                                      & Full &    -461.437753 &      -1.578592 &      -1.652826 &      -0.073256 &      -0.065721 &      -0.073716 \\ \hline
12    & Pyrazine dimer                       & F1   &    -262.702440 &      -0.835880 &      -0.862195 &      -0.038626 &      -0.034562 &      -0.038549 \\
      &                                      & F2   &    -262.702461 &      -0.835842 &      -0.862162 &      -0.038618 &      -0.034555 &      -0.038541 \\
      &                                      & Full &    -525.400639 &      -1.686240 &      -1.733417 &      -0.079039 &      -0.070599 &      -0.078493 \\ \hline
14    & Indolebenzene complex stack          & F1   &    -361.497850 &      -1.205004 &      -1.245524 &      -0.056259 &      -0.050375 &      -0.056045 \\
      &                                      & F2   &    -230.722148 &      -0.782743 &      -0.822212 &      -0.035828 &      -0.032199 &      -0.036226 \\
      &                                      & Full &    -592.211534 &      -2.007890 &      -2.080529 &      -0.094617 &      -0.084658 &      -0.094258 \\ \hline
16    & Etheneethyne complex                 & F1   &     -78.039902 &      -0.274863 &      -0.305019 &      -0.009799 &      -0.009002 &      -0.010430 \\
      &                                      & F2   &     -76.825504 &      -0.256064 &      -0.273040 &      -0.011148 &      -0.010133 &      -0.011540 \\
      &                                      & Full &    -154.866735 &      -0.532540 &      -0.579048 &      -0.021214 &      -0.019372 &      -0.022218 \\ \hline
17    & Benzenewater complex                 & F1   &    -230.722144 &      -0.782775 &      -0.822245 &      -0.035841 &      -0.032210 &      -0.036238 \\
      &                                      & F2   &     -76.026578 &      -0.202033 &      -0.211581 &      -0.003066 &      -0.002859 &      -0.003230 \\
      &                                      & Full &    -306.751679 &      -0.987721 &      -1.035750 &      -0.039306 &      -0.035417 &      -0.039820 \\ \hline
18    & Benzeneammonia complex               & F1   &    -230.722166 &      -0.782716 &      -0.822185 &      -0.035825 &      -0.032196 &      -0.036222 \\
      &                                      & F2   &     -56.195676 &      -0.186468 &      -0.202699 &      -0.003821 &      -0.003548 &      -0.004091 \\
      &                                      & Full &    -286.918481 &      -0.972582 &      -1.027292 &      -0.040078 &      -0.036121 &      -0.040699 \\ \hline
19    & BenzeneHCN complex                   & F1   &    -230.722122 &      -0.782834 &      -0.822305 &      -0.035856 &      -0.032224 &      -0.036253 \\
      &                                      & F2   &     -92.881359 &      -0.287354 &      -0.295666 &      -0.012245 &      -0.011064 &      -0.012427 \\
      &                                      & Full &    -323.607430 &      -1.074022 &      -1.120179 &      -0.048653 &      -0.043758 &      -0.049131 \\ \hline
20    & Benzene dimer TS                     & F1   &    -230.722171 &      -0.782715 &      -0.822184 &      -0.035823 &      -0.032194 &      -0.036221 \\
      &                                      & F2   &    -230.722165 &      -0.782736 &      -0.822205 &      -0.035829 &      -0.032200 &      -0.036227 \\
      &                                      & Full &    -461.443197 &      -1.572462 &      -1.649192 &      -0.072557 &      -0.065161 &      -0.073207 \\ \hline
\label{tab:s22}\\
\end{longtable}

\newpage

\newpage
\section{The plane wave based workflow}
\label{sec:planewave}

In this section, the workflow to calculate interaction energies of large molecules in a plane wave basis under periodic boundary conditions is described.
All calculations are performed with the Vienna Ab-Initio Simulation Package (VASP)~\cite{Efficiency.of.aKresse1996}  and the \texttt{Cc4s}~\cite{Gruber2018} code.

\begin{enumerate}

\item 
A fixed box size and a plane-wave basis set size are chosen.
The scheme is repeated for increasing box sizes to reach the infinite box size limit corresponding
to the isolated molecule in the gas phase.
The plane-wave basis set size was set via an energy cutoff of $700\,\text{eV}$ (\texttt{ENCUT} flag in VASP).
This choice resulted from a careful convergence test of the direct-MP2 correlation energy of the coronene dimer, achieving an accuracy well below 0.1 kcal/mol for a fixed box size.

\item 
The Hartree-Fock ground state is calculated using the given setting. Both the occupied as well as all unoccupied orbitals and orbital energies are stored.

\item 
Approximate natural orbitals at the MP2 level are calculated, as outlined in Ref.~\cite{Gruneis2011}.
Natural orbitals are the eigenvectors of the one-electron reduced density matrix.
The corresponding eigenvalues are called occupation numbers.
Ordered by their occupation number, we truncated and recanonicalized the natural orbital basis
by choosing a ratio $N_v/N_o$, where $N_o$ is the number of occupied orbitals in the system and $N_v$ is the number of chosen natural orbitals.
The natural orbitals provide a basis which allows for a much more rapid convergence of the correlation energy with respect to $N_v$.
\item 
The MP2 energy is calculated in the CBS limit using the natural orbitals with $N_v/N_o = 200$.
This is necessary for basis set correction schemes to estimate the CBS limit of the CCSD and (T) energies.
The basis set correction scheme, called focal point correction, is described in Ref. \cite{Irmler2021}.

\item 
To prepare the coupled cluster calculations a basis of $N_v/N_o = 15$ is chosen.
All Coulomb integrals, $V^{pq}_{sr}$, needed by coupled cluster theory are computed using the expression
\begin{equation}
V^{pq}_{sr}=\sum_{F=1}^{N_F}{\Gamma^*}_s^{pF}{\Gamma}_{rF}^{q},
\end{equation}
where $p,q,r,s$ refer to occupied or virtual orbital indices.
$F$ denotes an auxiliary basis functions, obtained
by a singular value decomposition outlined in Ref.~\cite{Hummel2017}.
Due to the large vacuum in the simulation cells, significant
reductions of the auxiliary basis set size are possible without
compromising the precision of computed correlation energies.
The correlation energies are converged to within meV with respect to the size of the optimized auxiliary basis set.

\item 
The final coupled cluster calculations at the level of CCSD, CCSD(T), and CCSD(cT) are performed with our high-performance code called \texttt{Cc4s}.
We employed up to 50 compute nodes with 128 cores each to run our massive computational parallelization approach.

\end{enumerate}

\subsection*{Benzene dimer (parallel displaced)}

\begin{figure}
  \centering
    \subfigure[\label{fig:benzene-no-convergence}]{\includegraphics[]{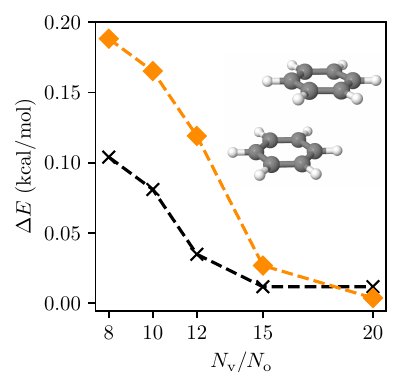}}
    \subfigure[\label{fig:benzene-vol-convergence}]{\includegraphics[]{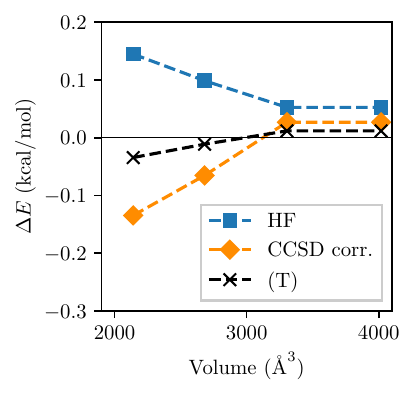}}
  \caption{%
    \textbf{Convergence behavior of the interaction energy for the benzene
    parallel displaced dimer system. }
    $\Delta E$ is the difference between our plane wave based approach and the reference results from Gaussian basis calculations extrapolated
to the CBS limit.
    \textbf{a,} shows the convergence with respect to the number of natural orbitals
    for a fixed volume of about $4018 \,\text\AA^3$.
    \textbf{b,} displays the convergence with respect to the volume of the simulation
    cell with a fixed number of natural orbitals of $N_\mathrm{v}/N_\mathrm{o}=15$.
  }
\end{figure}

We demonstrate that our plane wave basis approach works reliable for the study of noncovalent
interactions between molecules and combines the best of two worlds: compactness and systematic improvability
without linear dependencies.
To this end we discuss the computed interaction energy of the parallel displaced benzene dimer on the level of CCSD(T) theory and compare to results from basis set converged Gaussian calculations.
The total CCSD(T) energy is composed of three terms, the HF total energy,
the CCSD correlation energy and the perturbative triples contribution
which we denote as (T).
Fig.~\hyperref[fig:benzene-no-convergence]{S1(a)}
depicts the convergence of the CCSD and (T) correlation energy contributions to the computed CCSD(T) interaction energy for a fixed box size with respect to the number of basis functions (natural
orbitals) per occupied state ($N_\mathrm{v}/N_\mathrm{o}$).
We include a recently introduced correction to accelerate the convergence of correlation energies
to the complete basis set limit (CBS)~\cite{Irmler2021}.
Our findings show that a basis set size of $N_\mathrm{v}/N_\mathrm{o}=15$ suffices to achieve convergence to within a fraction of
$0.1\,\text{kcal/mol}$.
We employ this basis set to compute the CCSD(T) interaction energies and its Hartree--Fock (HF), CCSD and
(T) correlation energy contributions for different simulation cell sizes.
Fig.~\hyperref[fig:benzene-vol-convergence]{S1(b)} shows that these contributions converge rapidly.
Our fully converged estimate of the CCSD(T) interaction energy for the parallel displaced benzene dimer is $-2.62\,\text{kcal/mol}$,
which is in excellent agreement with results obtained using Gaussian basis sets of $-2.70\,\text{kcal/mol}$.

\subsection*{Coronene dimer (parallel displaced)}

\begin{figure}
  \centering
  \includegraphics[width=0.4\textwidth]{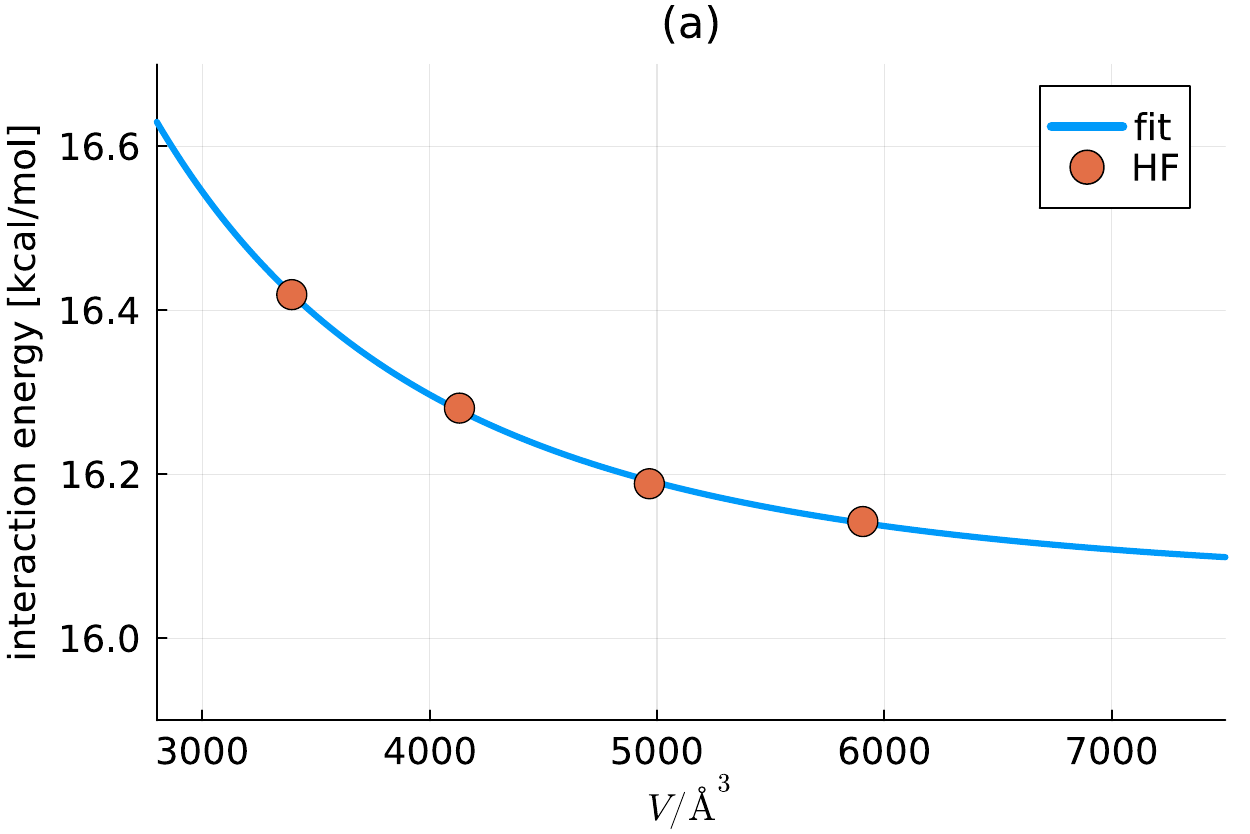}\\
  \includegraphics[width=0.4\textwidth]{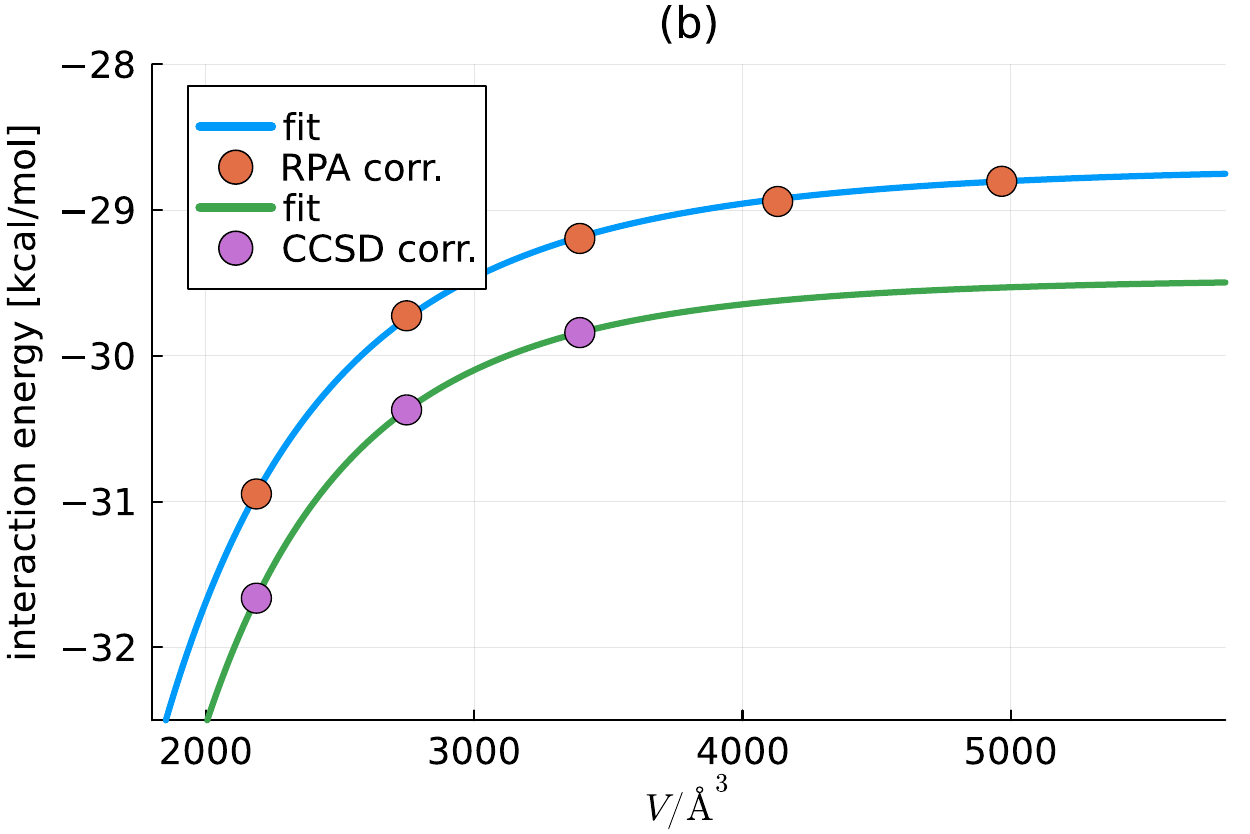}
  \includegraphics[width=0.4\textwidth]{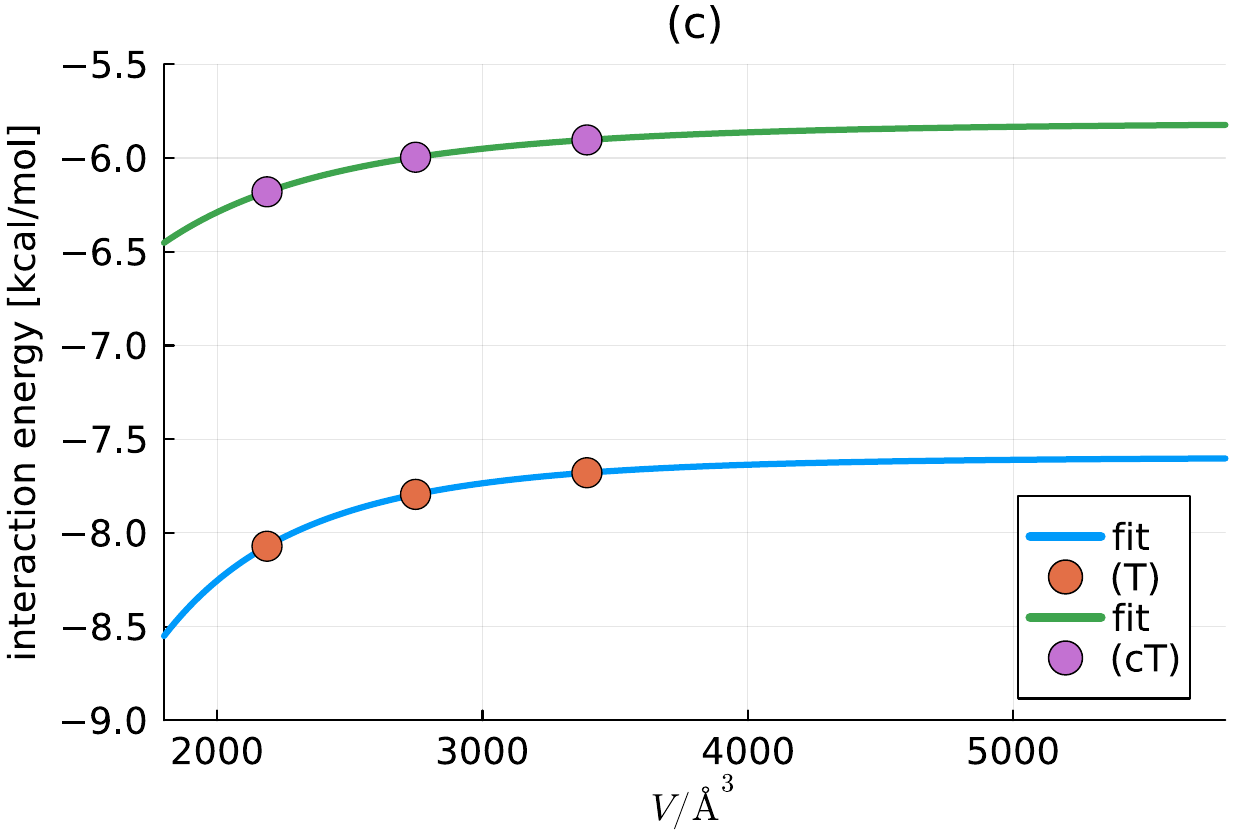}
  \caption{%
    Box size dependence of the interaction energy of the coronene dimer for (a) HF, (b) CCSD, and the (c) triples.
    The CCSD and triples contribution was calculated with a basis set of $N_v/N_o = 15$ and $N_v/N_o = 12$, respectively.
  }
  \label{fig:Vconv}
\end{figure}

\begin{figure}
  \centering
  \includegraphics[width=0.4\textwidth]{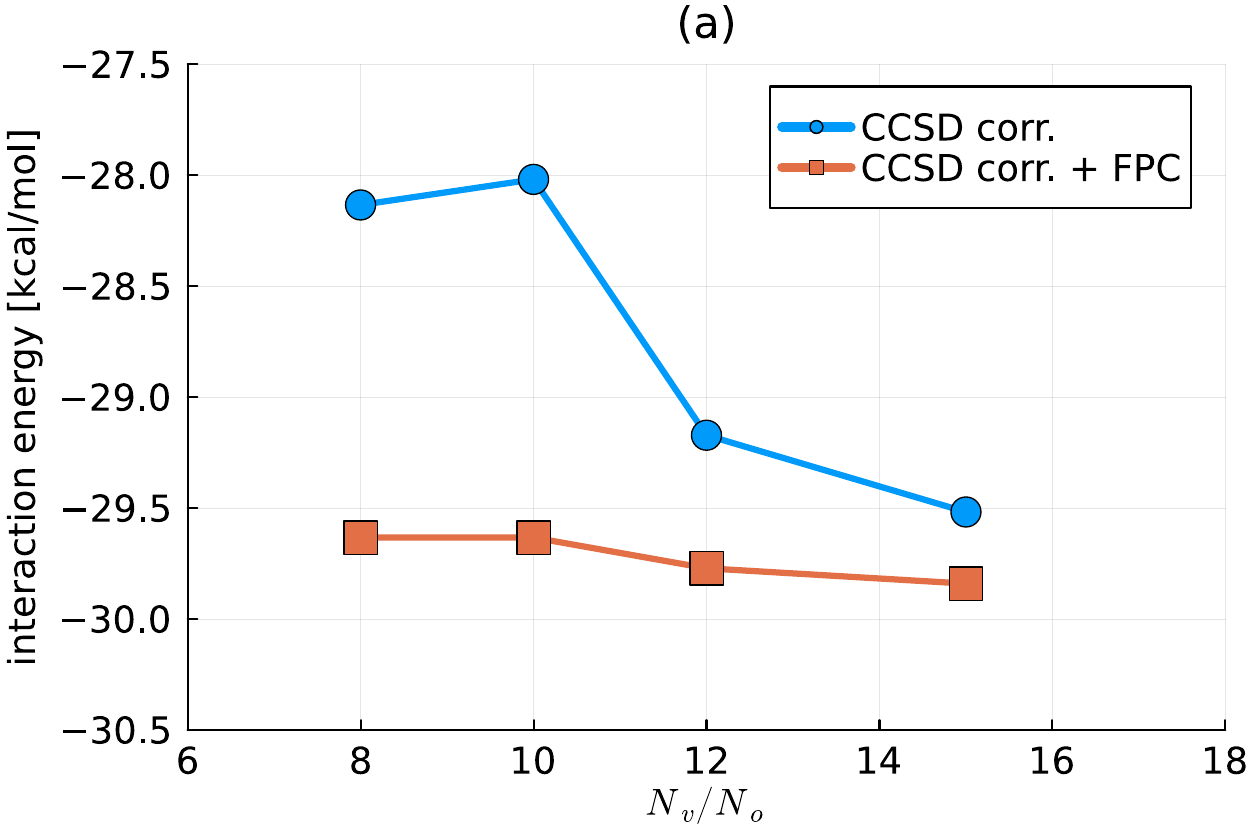}
  \includegraphics[width=0.4\textwidth]{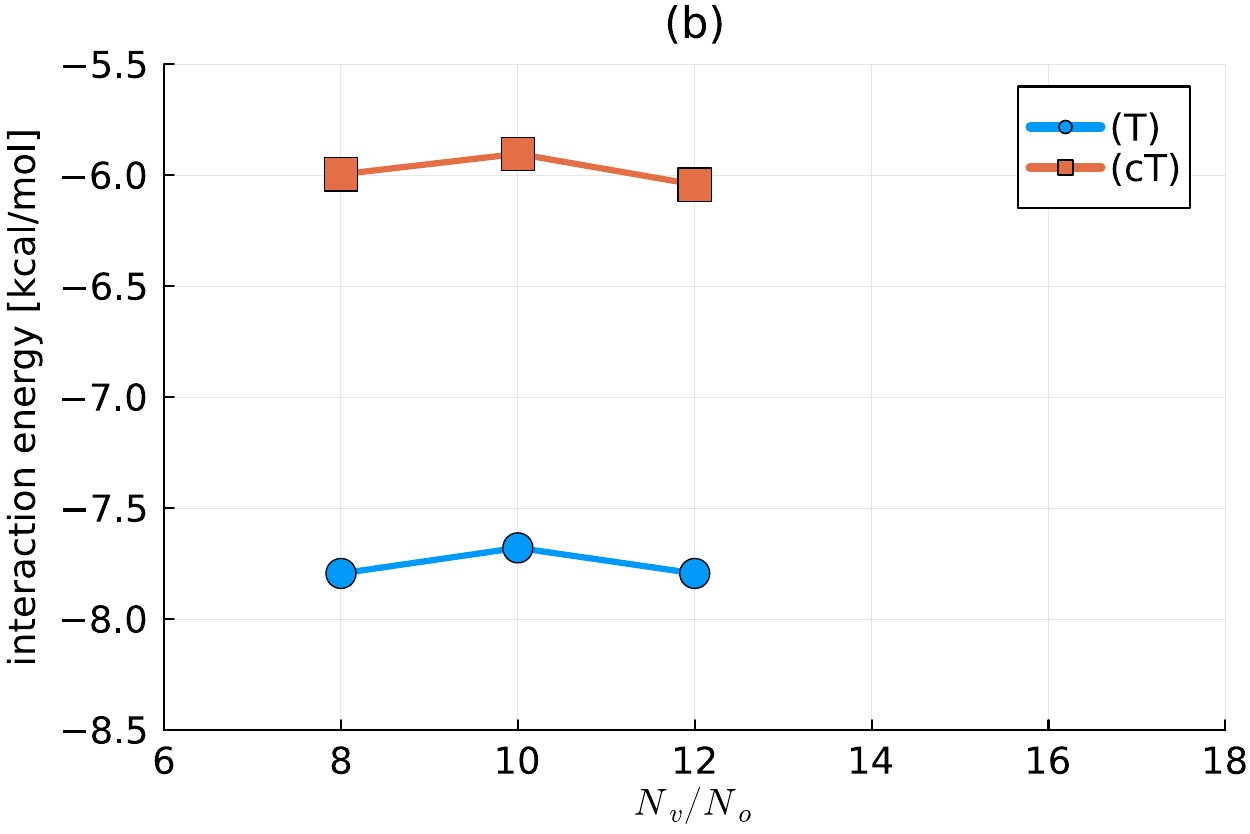}
  \caption{%
    Basis set convergence of the correlation contribution of the interaction energy of the coronene dimer at the level of (a) CCSD and the (b) triples.
    A box size of $3375 \,\text\AA^3$ was considered.
    FPC denotes a basis set correction scheme cited in the text.
  }
  \label{fig:NOconv}
\end{figure}

Using the example of the coronene dimer (C2C2PD), Fig.~\hyperref[fig:Vconv]{S2} shows the dependence of the interaction energy on the box size.
The interaction energy exhibits an exponential convergence of the form $a + b\cdot \text e^{-c V^{1/3}}$, where $V$ is the volume of the box.
This behavior holds for both the HF and correlation contributions.
The reliability of this extrapolation law is supported by RPA (random phase approximation) calculations of the correlation energy for volumes up to $\sim 5000 \text{\AA}^3$.
This allows us to converge the interaction energy with a remaining uncertainty of less than 0.5 kcal/mol.

The basis set dependence of the interaction energy is shown in Fig.~\hyperref[fig:NOconv]{S2}.
The additional focal point correction~\cite{Irmler2021} dramatically reduces the basis set error of the CCSD energy and allows us to consider $N_v/N_o = 15$ as a very good approximation to the complete basis set limit.
The triples contributions (T) and (cT) are corrected by rescaling the finite basis set result with a factor estimated on the level of MP2 theory as outlined in Ref~\cite{Knizia2009}.
Final CBS estimates of interaction energies for three systems in the L7 test set are provided in Table~\hyperref[tab:vasp]{S \RNum 3}.
\begin{table}[H]
\centering
\caption{%
  Results  of interaction energies in kcal/mol for three molecules from the L7 test set. CBS estimates obtained using the plane wave workflow as described in the text.
  LNO-CCSD(T) from GTO from literature are provided for comparison.
}
\begin{tabular}{lccccc|cc}
 \toprule
 System                      &   HF  & MP2 corr. & CCSD corr. & (T)     & (cT) & CCSD(T) & LNO-CCSD(T)~\cite{Hamdani2021} \\
 \midrule
 GGG                         &  7.725 & -12.245 & -7.771    & -1.384  & -1.130 & -1.430 & -2.1 $\pm$ 0.2 \\
 GCGC                        & 12.407 & -31.847 & -21.193   & -4.035  & -3.343 & -12.821 & -13.6 $\pm$ 0.4\\
 C2C2PD                      & 16.096 & -54.561 & -29.471   & -7.702  & -5.949 & -21.077 & -20.6 $\pm$ 0.6 \\
 \bottomrule
\end{tabular}
\label{tab:vasp}\\
\end{table}

\section{Estimating CCSD(cT)-fit and its uncertainty}
\label{sec:fit}

As presented in Fig. 4 in the article, a linear trend can be observed, when
plotting the ratio of (T) and (cT) against the ratio of the MP2 and CCSD
correlation contribution of the interaction energies. The corresponding (T),
(cT), MP2 as well as CCSD correlation energy contributions can be found in
Table~\hyperref[tab:ints]{S \RNum 4}.

A linear fit
\begin{equation}
\frac{(\text{T})}{(\text{cT})} = a + b \cdot  \frac{\text{MP2 corr.}}{\text{CCSD corr.}} \;,
\label{eq:fit}
\end{equation}
gives $a=0.7764$ and $b=0.2780$ with a standard deviation of the residuals of
$\sigma = 0.0097$.  Using this linear relationship, we derive the estimate
CCSD(cT)-fit for the interaction energy of large molecules via
\begin{equation}
\text{CCSD(cT)-fit} = \text{LNO-CCSD} + \frac{1}{X}\cdot\text{LNO-(T)} \;,
\end{equation}
where $X = a + b \cdot Y$ and $Y$ is the ratio of the MP2 and CCSD correlation
contribution obtained from the LNO coupled cluster approach.

This procedure allows us to calculate CCSD(cT) estimates for the large
molecules presented in the work of Hamdani \textit{et al.}~\cite{Hamdani2021}.
Therefore we calculate $\Delta = \text{CCSD(T)} - \text{CCSD(cT)}$ using the
LNO-CCSD(T) results from Table~\hyperref[tab:lno]{S \RNum 5} together with
Eq.~\ref{eq:fit}. This energy difference $\Delta$ is subtracted from the the
well-converged LNO-CCSD(T) estimates provided in Ref.~\cite{Hamdani2021}.

We estimate the uncertainty of the CCSD(cT)-fit interaction energy as the sum
of the LNO-CCSD(T) uncertainty provided in Ref.~\cite{Hamdani2021} and an
uncertainty from the fit.  The former is a consequence of the tightness
parameters controlling the local approximation.  The latter can simply be
calculated via the standard deviation of the residuals $\sigma$, which provides
an error estimate for $X$.  In correspondence with the uncertainty of the DMC
results, which take $2\sigma$, our corresponding error estimate for $1/X$ is
thus given by $2\sigma/X^2$.  Hence, the uncertainty measure for the fit
depends on the considered molecular system but roughly takes the value of
$2\sigma/X^2\approx 0.025$ for all considered cases.  Finally. this leads to an
error estimate of
\begin{equation}
\delta\big( \text{CCSD(cT)-fit} \big) = \delta\big( \text{LNO-CCSD(T)} \big) + \left| \frac{2\sigma}{X^2}\cdot\text{LNO-(T)} \right| \;.
\end{equation}
In fact, we make the simplification that $\delta\big( \text{LNO-CCSD(T)} \big)$
and $\delta\big( \text{LNO-CCSD} \big)$ are similar, as only the former is
provided in Ref.~\cite{Hamdani2021}.

\begin{table}[H]
\centering
\caption{%
Interaction energies in kcal/mol of a set of dispersion-dominated complexes
from the S22, L7 and S66 benchmark datasets.  Systems from the S22 test set are
taken from Table~\hyperref[tab:s22]{S \RNum 2} and are calculated using cc-pVDZ
basis sets.  Systems from the S66 are obtained from [34] extrapolation and were
taken from Table~\hyperref[tab:mrcc-mp2]{S \RNum 1}.  Results for the three
molecules from the L7 test set are obtained from plane wave calculations and
were taken from Table~\hyperref[tab:vasp]{S \RNum 3}.
}
\begin{tabular}{lcccc}
 \toprule
 System                      &  MP2 corr. & CCSD corr. & (T)     & (cT)  \\
 \midrule
 Methane dimer               &  -0.619   & -0.559    & -0.064  & -0.059  \\
 Ethene dimer                &  -1.797   & -1.368    & -0.244  & -0.218  \\
 Benzene-Methane complex     &  -2.424   & -1.810    & -0.299  & -0.260  \\
 Benzene dimer PD            &  -8.312   & -5.361    & -1.025  & -0.849  \\
 Pyrazine dimer              &  -9.111   & -5.685    & -1.126  & -0.930  \\
 Indolebenzene complex stack &  -12.640  & -8.028    & -1.587  & -1.308  \\
 \hline
 Pyridine-Pyridine PD        &  -9.339   & -5.794    & -1.310  & -1.080  \\
 Pyridine-Pyridine TS        &  -5.258   & -3.597    & -0.762  & -0.642  \\
 Benzene-Pyridine PD         &  -9.052   & -5.651    & -1.282  & -1.058  \\
 Benzene-Pyridine TS         &  -5.119   & -3.488    & -0.735  & -0.618  \\
 Pyridine-Uracil PD          &  -10.701  & -7.110    & -1.637  & -1.362  \\
 Benzene-Benzene PD          &  -8.665   & -5.423    & -1.234  & -1.019  \\
 Benzene-Benzene TS          &  -5.194   & -3.518    & -0.748  & -0.628  \\
 Uracil-Uracil PD            &  -11.525  & -8.142    & -1.972  & -1.669  \\
 Benzene-Uracil PD           &  -10.982  & -7.334    & -1.670  & -1.394  \\
\hline
 GGG                         &  -12.245  & -7.771    & -1.384  & -1.130  \\
 GCGC                        &  -31.847  & -21.193   & -4.035  & -3.343  \\
 C2C2PD                      &  -54.561  & -29.471   & -7.702  & -5.949  \\
 \bottomrule
\end{tabular}
\label{tab:ints}\\
\end{table}

\renewcommand*{\arraystretch}{1.1}
\begin{longtable}[H]{rcccc|cc|c}
\caption{%
Results for the L7 molecules and C$_{60}$[6]CPPA using the LNO-CCSD(T)
algorithm in \texttt{MRCC}.  We use aug-cc-pVTZ basis sets and employ
counterpoise correction. In all calculations we use as LNO threshold the
keyword Tight. Except for the C$_{60}$[6]CPPA molecule where we have employed
the threshold Normal. CCSD and (T) correlation energy contain each half of the
MP2 correction originating from weak pairs.  In addition to the individual
energy contributions we show our LNO-CCSD(T) results in comparison with the CBS
estimates published in Ref.~\cite{Hamdani2021}.  $\Delta$ is the estimated
difference between (T) and (cT) based on the described fitting procedure.
} \\
\toprule
System   &    HF & MP2 corr.&   CCSD corr. &  (T) & LNO-CCSD(T)   & LNO-CCSD(T)~\cite{Hamdani2021} & $\Delta$\\
 \midrule
 \endhead
 \bottomrule
 \endfoot
GGG &
   8.160 & -12.365 &   -8.380 &  -2.142 &  -2.362 &  -2.100 & -0.337\\
GCGC &
  12.317 & -30.221 &  -20.840 &  -5.305 & -13.828 & -13.600 & -0.808\\
C2C2PD &
  15.974 & -53.177 &  -29.888 &  -8.091 & -22.005  & -20.600 & -1.725\\
C3A &
   9.291 & -35.507 &  -20.709 &  -5.848 & -17.266 & -16.500 & -1.181\\
PHE &
 -13.644 & -11.463 &   -8.109 &  -2.641 & -24.394 & -25.400 & -0.383\\
C3GC &
  17.486 & -61.393 &  -36.794 & -10.305 & -29.614 & -28.700 & -1.996 \\
C$_{60}$@[6]CPPA &
   56.349 &  -141.173&  -78.777 & -28.508 & -50.936  & -41.700 & -6.142
\label{tab:lno}
\end{longtable}

\bibliography{refs}